\documentclass[preprint]{aastex6}
\usepackage{amsmath}
\usepackage{color}
\usepackage{comment}
\usepackage{url}
\usepackage{comment}

\usepackage{graphicx}

\def\commenta{$^*$}
\def\commentb{$^\dagger$}
\def\commentc{$^\ddagger$}
\def\commentd{$^\S$}
\def\commente{$^\|$}
\def\commentf{$^\#$}
\def\commentg{$^{**}$}

\begin{document}
\shorttitle{Lifetimes and Emergence/Decay Rates of Star Spots on Solar-type Stars}
\shortauthors{Namekata et al.}


\title{Lifetimes and Emergence/Decay Rates of Star Spots on Solar-type Stars Estimated by \textit{Kepler} Data in Comparison with Those of Sunspots}

%\begin{comment}

%% Use \author, \affil, plus the \and command to format author and affiliation 
%% information.  If done correctly the peer review system will be able to
%% automatically put the author and affiliation information from the manuscript
%% and save the corresponding author the trouble of entering it by hand.
%%
%% The \affil should be used to document primary affiliations and the
%% \altaffil should be used for secondary affiliations, titles, or email.

%% Authors with the same affiliation can be grouped in a single
%\author{Greg J. Schwarz\altaffilmark{1,2} and August Muench\altaffilmark{1}}
%\affil{American Astronomical Society \\
%2000 Florida Ave., NW, Suite 300 \\
%Washington, DC 20009-1231, USA}

%\author{Butler Burton\altaffilmark{3}}
%\affil{National Radio Astronomy Observatory}

%\affil{TeXnology Inc}

%\author{Julie Steffen\altaffilmark{4}}
%\affil{American Astronomical Society \\
%2000 Florida Ave., NW, Suite 300 \\
%Washington, DC 20009-1231, USA}

\author{Kosuke Namekata\altaffilmark{1}}
\author{Hiroyuki Maehara\altaffilmark{2, 3} }
\author{Yuta Notsu\altaffilmark{1} }
\author{Shin Toriumi\altaffilmark{4} }
\author{Hisashi Hayakawa\altaffilmark{5}}
\author{Kai Ikuta\altaffilmark{1}}
\author{Shota Notsu\altaffilmark{1} }
\author{Satoshi Honda\altaffilmark{6}}
\author{Daisaku Nogami\altaffilmark{1} }
%% Use the \and command so offset the last author.
\and

\author{Kazunari Shibata\altaffilmark{7}}


\altaffiltext{1}{Department of Astronomy, Kyoto University, Kitashirakawa-Oiwake-cho, Sakyo, Kyoto 606-8502, Japan; namekata@kusastro.kyoto-u.ac.jp}

\altaffiltext{2}{Okayama Branch Office, Subaru Telescope, National Astronomical Observatory of Japan, NINS, Kamogata, Asakuchi, Okayama 719-0232, Japan}

\altaffiltext{3}{Okayama Observatory, Kyoto University, 3037-5 Honjo, Kamogata, Asakuchi, Okayama 719-0232, Japan.}

\altaffiltext{4}{National Astronomical Observatory of Japan, NINS, Mitaka, Tokyo 181-8588, Japan.}

\altaffiltext{5}{Graduate School of Letters, Osaka University, Toyonaka, Osaka 560-0043, Japan.}

\altaffiltext{6}{Nishi-Harima Astronomical Observatory, Center for Astronomy, University of Hyogo, Sayo, Sayo, Hyogo 679-5313, Japan.}

\altaffiltext{7}{Astronomical Observatory, Kyoto University, Kitashirakawa-Oiwake-cho, Sakyo, Kyoto 606-8502, Japan.}

\begin{abstract}
%太陽型星も含めた種系列恒星は、高い磁気活動を示すことがあるが、その磁気活動の起源はよくわかっていない
%ケプラー衛星によって

Active solar-type stars show large quasi-periodic brightness variations caused by stellar rotations with star spots, and the amplitude changes as the spots emerge and decay.
The Kepler data are suitable for investigations on the emergence and decay processes of star spots, which are important to understand underlying stellar dynamo and stellar flares.
In this study, we measured temporal evolutions of star spot area with Kepler data by tracing local minima of the light curves.
In this analysis, we extracted temporal evolutions of star spots showing clear emergence and decay without being disturbed by stellar differential rotations.
We applied this method to 5356 active solar-type stars observed by Kepler and obtained temporal evolutions of 56 individual star spots.
We calculated lifetimes, emergence and decay rates of the star spots from the obtained temporal evolutions of spot area.
As a result, we found that lifetimes ($T$) of star spots are ranging from 10 to 350 days when spot areas ($A$) are 0.1--2.3 percent of the solar hemisphere. 
We also compared them with sunspot lifetimes, and found that the lifetimes of star spots are much shorter than those extrapolated from an empirical relation of sunspots ($T\propto A$), while being consistent with other researches on star spot lifetimes.
The emerging and decay rates of star spots are typically $5 \times 10^{20}$ $\rm Mx\cdot h^{-1}$ ($8$ $\rm MSH\cdot h^{-1}$) with the area of 0.1--2.3 percent of the solar hemisphere and are mostly consistent with those expected from sunspots, which may indicate the same underlying processes.
\end{abstract}

%Recently, many superflares on solar-type stars were discovered by the Kepler Space Telescope.
%Such active stars are thought to have large star spots.
%The emerging and decay process of such large star spots are not well understood, but important for the understanding of superflare events as well as underlying stellar dynamo.


\section{Introduction} \label{sec:int}

Recent studies have shown that flaring activities and spot generations are seen not only on the Sun but also on other solar-type stars (G-type main-sequence stars). 
%These phenomena may bridge the knowledge between solar and stellar physics, which may share common underlying physical mechanisms.
These phenomena may share common underlying mechanisms, which therefore bridges the solar and stellar physics communities.
Interestingly, some of the solar-type stars show extremely high magnetic activities that are not expected from the 150-year solar observations \citep{2013JSWSC...3A..31C,2015LRSP...12....4H}.
They possess giantic spots with areas of 10,000-100,000 MSH \citep[millions of solar hemisphere, $10^{6}$ MSH = $2\pi R_{\odot}^2$ = 3.1$\times 10^{22}$ cm$^2$,][]{2017PASJ...69...41M}, which is by far larger than the largest sunspot \citep[6132 MSH, ][]{2013A&A...549A..66A,2017ApJ...834...56T,2017ApJ...850L..31H}, and they sometimes show extraordinary large flares called superflares \citep[$10^{34-36}$ erg,][]{2012Natur.485..478M,2013ApJS..209....5S}.
It is indicated that superflares occur by the same process as solar flares, i.e., through magnetic reconnection \citep{2013ApJ...771..127N,2015EP&S...67...59M,2016NatCo...711058K,2017PASJ...69....7N}.
%The mechanism to generate the large star spots has been focused on for the understanding of superflare events \citep[e.g.,][]{2013PASJ...65...49S}. %(Shibata et al. 2013).
However, little is known about the process to generate the magnetic fields of the spots.
Since the spots are visible proxies of magnetic field on the stellar surface \citep[e.g.,][]{2017LRSP...14....4B}, observations of the spot properties may provide a clue to the universal understanding of the solar and stellar dynamo \citep[e.g.,][]{2013PASJ...65...49S}.

In the case of sunspots, the magnetic fluxes are generally thought to emerge from the deep convection zone to the solar surface thanks to the magnetic buoyancy and convection
\citep[e.g.,][]{1955ApJ...121..491P,2014LRSP...11....3C}, and decay as soon as (even before) the sunspots are completely formed \citep{1981phss.conf....7M}.
%Complicatingly, the decay timescales do not match the dynamical timescale ($\tau_{\rm dyn}$ $\sim$ hours) or the magnetic diffusion timescale in the photosphere ($\tau_{\rm diff}$ $\sim$ years) \citep{2011LRSP....8....3R}, making it difficult to identify the decay mechanism.
As for the decay process, the granular motions can be a possible mechanism, though other processes such as surface flows \citep[moat/Evershed flows, e.g., ][]{2008ApJ...686.1447K}, subsurface convections and reconnections \citep{1999SoPh..188..331S} can also contribute to the spot decay.
It takes typically hours to days for the sunspot formations \citep[$<$ 5 days;][]{1993SoPh..148...85H}, while the spot decay spends longer periods, typically weeks to months \citep{2008SoPh..250..269H}.
Since the decay phase is longer than the emerging phase, sunspot lifetimes have been discussed with regard to the decay mechanism.
Spot decay rates ($dA/dt$) as a function of the spot area ($A$) have been best discussed so far.
There are two models, a linear decay law \citep[$dA/dt \propto$ constant;][]{1963BAICz..14...91B} and a parabolic decay law \citep[$dA/dt \propto A^{1/2}$;][]{1993A&A...274..521M}.
In the former case, a relation between the lifetime ($T$) and spot area ($A$) can be formulated into $T\propto A$, which is called "Gnevyshev-Waldmeier" (GW) law \citep{Gnevyshev,1955epds.book.....W}.
In the latter case, the lifetime-area relation can be expressed as $T\propto A^{1/2}$ \citep{1993A&A...274..521M}, while $T\propto A$ can be also derived by considering the maximum size dependency \citep{1997SoPh..176..249P}.
Observationally, lifetimes of sunspot groups are roughly consistent with the GW law, though the large scattering around the GW law can be seen \citep[e.g., ][]{2010SoPh..262..299H}.

Star spots are also universally observed on various kinds of stars, including solar-type stars \citep[see][for review]{2005LRSP....2....8B}.
Stars having large spots show large quasi-periodic brightness variations caused by stellar rotations \citep{2015PASJ...67...33N,2016NatCo...711058K}, which helps us investigate the star spot properties.
Although the temporal evolutions of star spots can be a clue to the understanding of the supply and dissipation of magnetic field on stellar surface, they have not been extensively investigated due to its difficulty in observation.
However, investigations on temporal evolutions of star spots may exert a huge impact on a variety of research fields because of the following reasons.
%(1) it can reveal basic plasma physics on star spots;
(1) It can be a tool to understand how the superflares are triggered, and enable us to estimate how long surrounding planets are exposed to the danger of flares and coronal mass ejections \citep[e.g.,][]{2016ApJ...833L...8T,2017ApJ...848...41L}.
(2) Estimating the diffusion coefficient of stellar surface would be helpful for numerical modelings on stellar dynamo.
(3) It can provide a constraint for the light curve modeling calculations to reconstruct surface intensity distributions, which are helpful for detections of exoplanet transits \citep{2017MNRAS.472.1618G}.
%%Rotationa period determinationも

In 1990s, several researches reported lifetimes of star spots on active young stars, cool stars (mainly M and K-type stars), and RS CVn-type stars (close binary stars) on the basis of ground-based observations \citep[e.g.,][]{1994IAPPP..55...51H,1995ApJS...97..513H,1994A&A...282..535S}.
They indicated that the lifetimes of star spots linearly increase with the spot area in the domain of small spots, while they decrease in the domain of large spots possibly because of differential rotations \citep{1995ApJS...97..513H}.
For the huge spots, the estimated lifetimes are quite long and sometimes exceed 2 years \citep{1994A&A...282..535S}.
Futheremor, the developments of Doppler Imaging techniques surprisingly revealed that active young stars have large polar spots which have persisted for more than decades \citep{1999A&A...347..212S,2012A&A...548A..95C}.
More recently, the light curve modelings have been carried out by many authors \citep[see,][for review]{2009A&ARv..17..251S} and an inversion modeling has beed developed \citep{2008AN....329..364S}, which may reveal the spot temporal evolutions although there may be more or less degeneracy in inversions.

In 2009, the Kepler satellite was launched to observe a huge amount of long-term stellar light curves ($\sim$4 years), and the high-sensitivity observations have enabled us to research the temporal evolutions of star spots on solar-type stars by using several methods.
\cite{2012A&A...543A.146F} performed a light curve modeling to reconstruct the spot evolutions for two solar-type stars, though their applications have been limited to short observational periods ($\sim$ 130 days),  which does not clearly cover the whole spot lifetimes.
\cite{2014ApJ...795...79B} and \cite{2015PhDT.......177D} estimated lifetimes and area of star spots on solar-type stars on the basis of the surface distributions reconstructed by the stellar brightness variation during exoplanet transits.
They showed that star spots with area of 10,000--100,000 MSH have much shorter lifetimes ($\sim$10--200 day) than those expected from the solar GW relation (1,000--10,000 day).
More recently, \cite{2017MNRAS.472.1618G} developed a method to derive an indicator to characterize the star spot lifetime on solar-type stars by applying the auto-correlation function to the Kepler light curves.
Although this is only a proxy of the typical spot decay time of the star, it enables a statistical study for large samples and interestingly shows a trend similar to that in \cite{2015PhDT.......177D}.
%The result of \cite{2017MNRAS.472.1618G} may not exactly correspond to lifetimes of single star spots but those of active longitude, 
%and \cite{2014ApJ...795...79B} and \cite{2015PhDT.......177D} analyzed only 2 stars having hot-jupiter planets.
What we still need to do for revealing the universal star spot physics are (1) more statistical analyses of the detailed temporal evolutions of individual star spots and (2) comparison with sunspots whose properties are well known.

In this study, we develop a method to measure temporal evolutions of star spots on solar-type stars by tracing local minima of the Kepler light curves.
This enables us to estimate not only lifetime--area relation but also emergence and decay rates of star spots.
In Section \ref{sec:ana}, we introduce our sample selection, method, and detection criteria.
We also describe how to calculate lifetimes, areas, and emergence and decay rates.
In Section \ref{sec:res}, we show several results of our analysis and comparisons with the solar data.
Finally, we discuss the results in Section \ref{sec:dis}.

\section{Data and Analysis Method} \label{sec:ana}
\subsection{Sample Selection}
It is expected that the spot emergence, its decay, and dynamo mechanism are all closely related to the stellar surface temperature and gravity \citep[e.g.,][]{1990sse..book.....K}.
In order to assess the diversity and similarity of the star spots by comparing them with the sunspots, we here selected solar-type stars (G-type main sequence stars) as target stars from the \textit{Kepler} data set on the basis of the stellar effective temperature ($T_{\rm eff}$) and surface gravity (log $g$) listed in Kepler Input Catalog \citep[\textit{Kepler} Data Release 25 Notes,][]{2016KeplerDR25}.
In this study, we defined solar-type stars with a criteria of 5000 K $<$ $T_{\rm eff}$ $<$ 6000 K and log$g$ $>$ 4.0.
For each star, we used all the available Kepler pre-search data conditioning (PDC) long-cadence (30 min) data \citep[\textit{Kepler} Data Release 25,][]{2016KeplerDR25}, in which instrumental effects are removed.

The active stars show quasi-periodic variations due to stellar rotations with large dark star spots,
which form local minima in the light curves corresponding to times when star spots are on the visible side of the stars.
The brightness variation amplitude corresponds to the spot area compared to stellar disk \citep[e.g.,][]{2013ApJ...771..127N,2015PASJ...67...33N}.
The idea of this study is that temporal evolutions of the star spots are measurable if we can trace the local minima in time series, as introduced in the following section.
In inactive stars, it is difficult to detect and trace the local minima because of the low signal to noise ratio.
We therefore only selected stars showing high magnetic activity with an additional criterion: amplitude of periodic variability taken from \cite{2014ApJS..211...24M} are above 1\%.
The 5356 active solar-type stars are finally selected as our target stars.

\subsection{Detection and Tracing of Local Minima}\label{sec:ana:trace}

\begin{figure}[htbp]
\begin{center}
\includegraphics[scale=0.6]{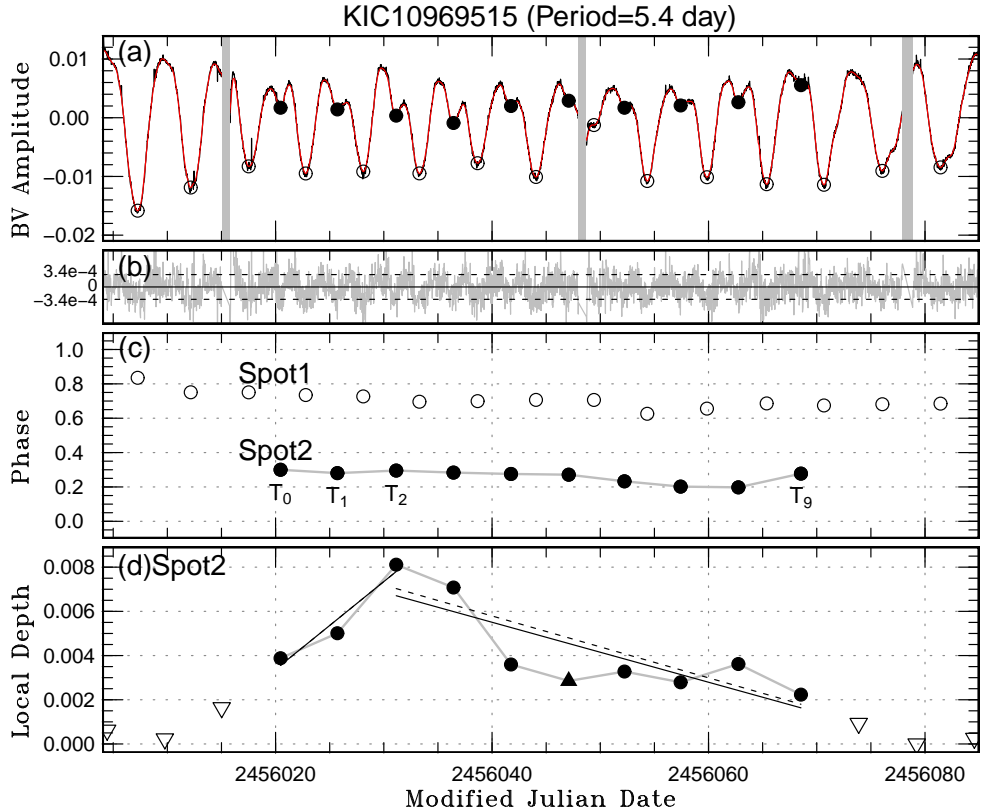}
\end{center}
\caption{An example of the temporal evolution of one star spot successfully measured in this study.
(a): Background black line is  the observed \textit{Kepler} light curves, and red line is the fitted one.
Vertical gray lines correspond to the observational gaps longer than 5 hours.
Filled circles are the local minima of a detected spots candidate, and the open circles are the local minima of another spot.
(b): The residual errors between the original Kepler light curve and fitted one. 
%Data are plotted from 3 rotational period before the emergence to 3 rotational period after the disappearance.
%Filled and open circles are detected local minima of the light curves, and filled ones are the spots we focused on.
(c): Phase-time diagram, where the vertical axis corresponds to phases of the local minima detected in the upper panels compared to the rotational period (cf., Carrington longitude).
Symbols are the same as the above.
Open circles correspond to another spot candidates, but they are not included in our catalog because their temporal evolutions are not well measurable.
(d): The temporal evolution of the depth of the local minima from the nearby local maximums of the spot candidates that we focus on.
The filled upward triangle is the local minima near the observational gaps.
The open downward triangles indicate the upper limit of the brightness variation before the spot emerge and after the spot disappear.
The black solid lines are fitted lines of the emergence and decay phase (for data unaffected by data gaps), and dashed lines are the same as solid one but fitted only for black circles.
Figures for all the other 55 spot candidates are available in the online version of the Journal.}
\label{fig:oc0}
\end{figure}

We used a simple method similar to that of \cite{1994IAPPP..55...51H} to measure the temporal evolutions of star spots.
In this method, each star spot can be identified by the repetition of the local minima over the rotational phases (see below Figure \ref{fig:oc0}).
The light curve of a rotating star with star spots shows several local minima when the spots are on the visible side (Figure \ref{fig:oc0}(a)).
The time separation of each local minimum corresponds to the rotational period.
For example, if a star has two large star spots at separated longitudes, the light curve exhibits two local minima during one rotational period, and the separation of the two local minima give the difference of the longitudes.
This difference makes it possible to identify longitudinally-separated star spots.
In the time-phase diagram of local minima  (Figure \ref{fig:oc0}(c)), an individual star spot is distinguishable as a common straight line (e.g., the gray line in  Figure \ref{fig:oc0}(c)).
This is how the individual star spots are identified and their temporal evolutions are measured from the light curves (Figure \ref{fig:oc0}(d)).
This method is our basic idea to discuss the temporal evolution of star spots, and has been applied to the ground-based observation of young stars, cool stars, and RS-CVn stars \citep[e.g.,][]{1995ApJS...97..513H}. %%%%%%

However, this method contains a problem caused by the stellar differential rotation.
If two spots are located at the different latitudes, the time separation of two local minima changes in time because of the differential rotation \citep[see, e.g.][]{1992A&A...259..183S}.
The differential rotation finally makes the two local minima combined to one.
This difficulty prevents us from tracing the whole time evolutions of the identical star spots from the appearance to disappearance.
Therefore, this method cannot distinguish whether the spot disappears or combines with other spots at the same longitude.
Moreover, changes in relative longitudes of the spots lead to changes in depths of local minima, which makes it difficult to estimate variation rates of the spot area \citep[see][]{2013ApJ...771..127N}.

To overcome these difficulties, we introduce the following conditions: 
we focus on light curves that have a pair of spots (1)  rotating  with a common period and (2) located on the reverse rotational phase (i.e. a longitude separation of the two spots of approximately 180 degree).
As for the pair of spots satisfying the condition (1), the absolute values of spot latitudes are considered the same.
When a light curve satisfies the conditions (1) and (2), the local minima can be traced without being disturbed by the differential rotation as well as brightness variations of the other spots.
Although this method can contain some selection biases, the simplicity enables an application to a huge amount of the \textit{Kepler} dataset.

\begin{figure}[htbp]
\begin{center}
\includegraphics[scale=0.32]{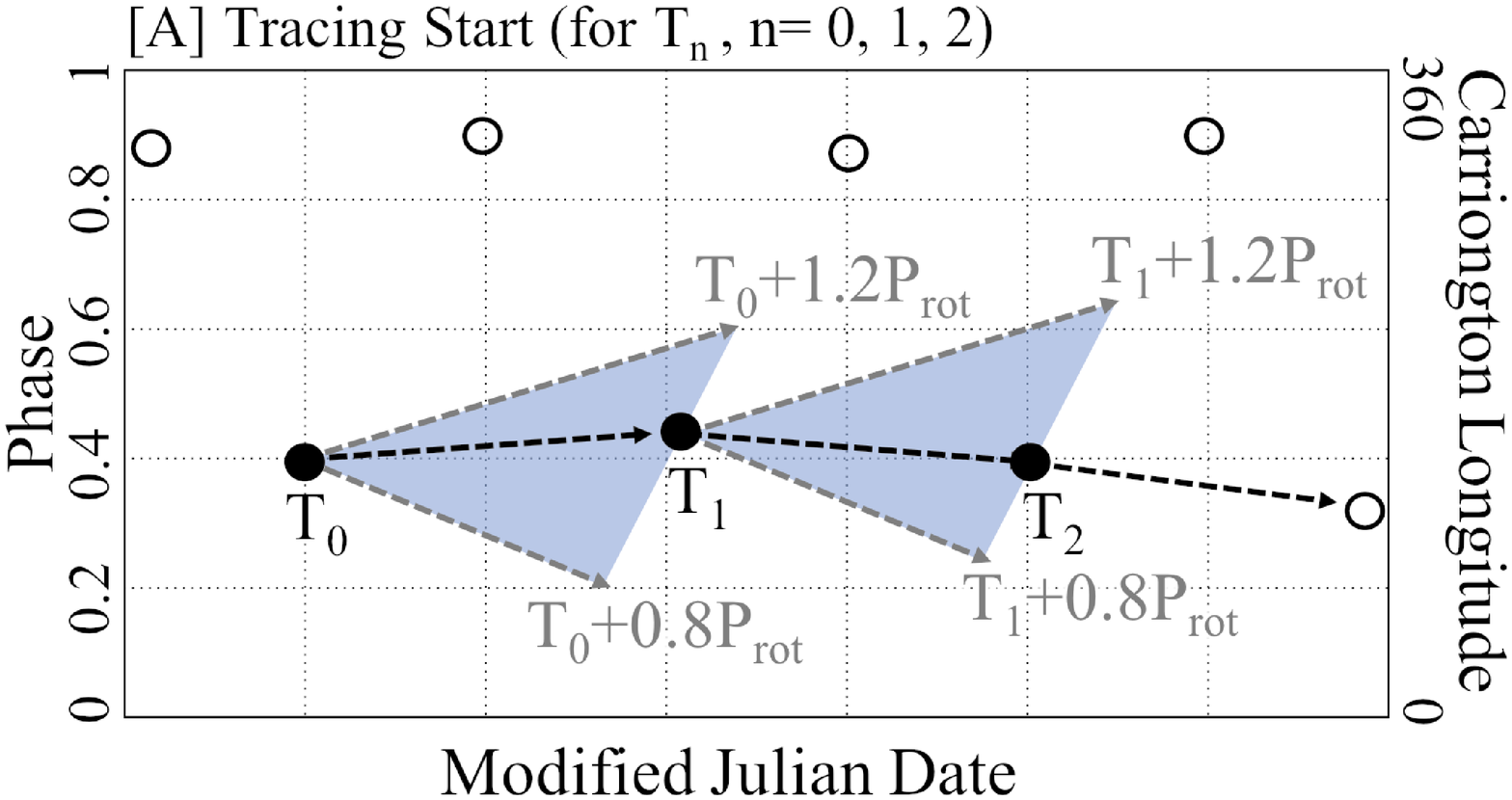}
\includegraphics[scale=0.32]{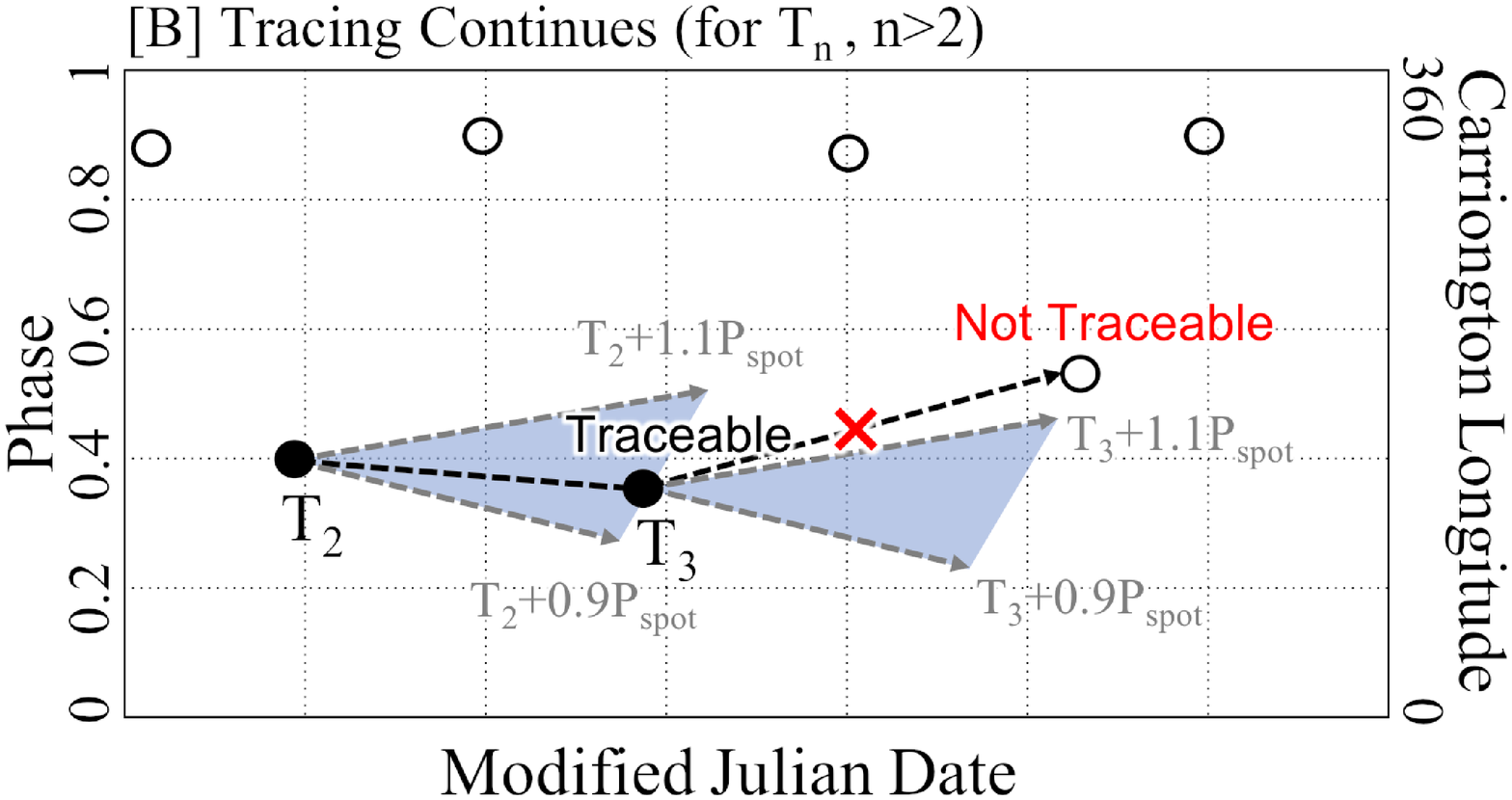}
\includegraphics[scale=0.32]{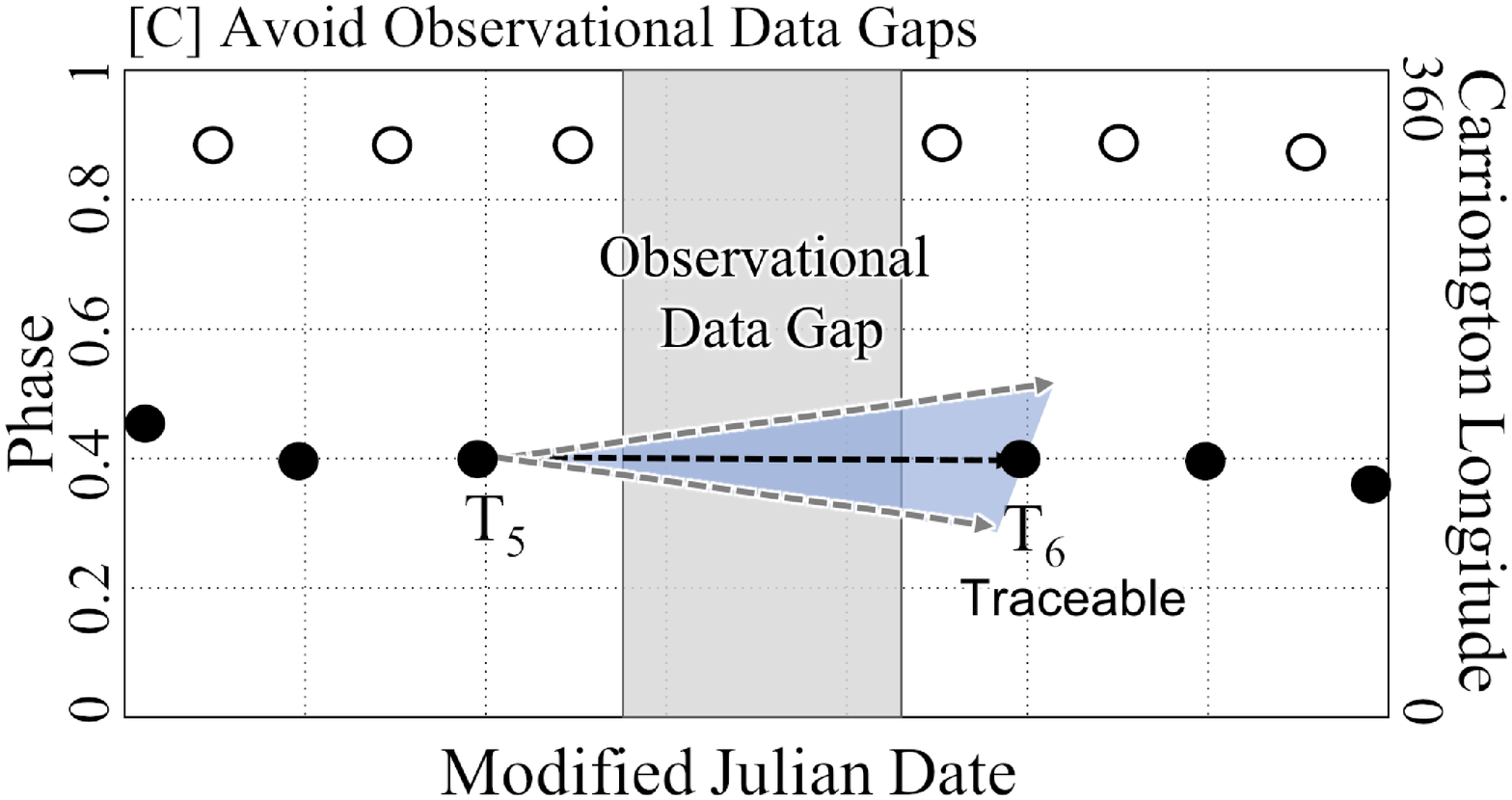}
\includegraphics[scale=0.32]{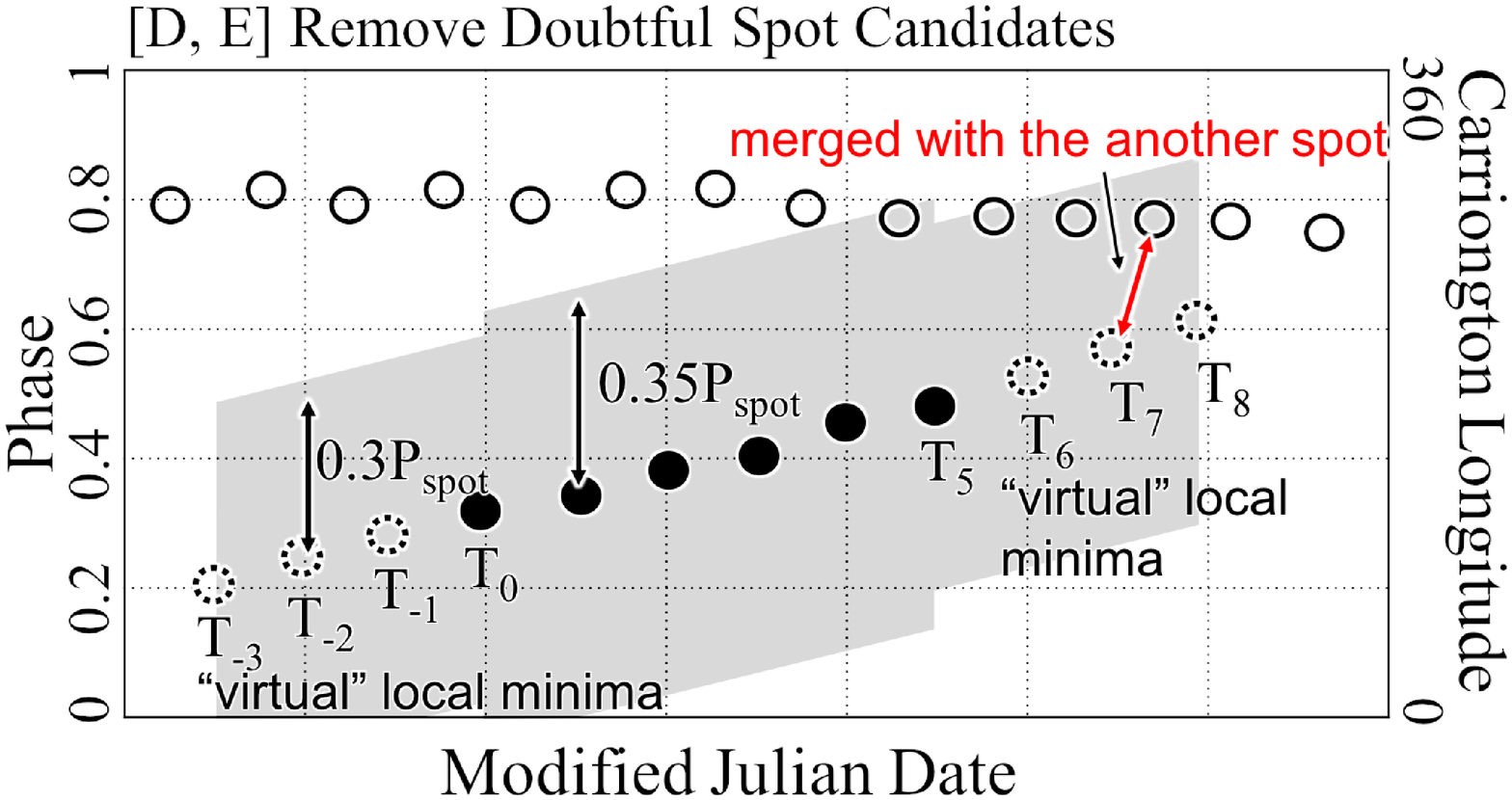}
\end{center}
\caption{Schematic pictures to explain the method ([A] $\sim$ [E]) of tracing the local minima described in Section \ref{sec:ana:trace}.
All panels correspond to the phase-time diagram of local minima, which is similar to Figure \ref{fig:oc0}(c). 
Each panel describe the following; [A] explains how to start the tracing and identification of spots, [B] how to continue the tracing after the identifications, [C] how to avoid the observational data gaps, and [D, E] how to remove the spot candidates that can merge with other spots in longitudes which are originally located at the separated longitudes. If there are other local minima in the gray region, the spot candidates are removed in our algorithm.}
\label{fig:analysis}
\end{figure}

Based on the above idea, we developed an algorithm to automatically detect such star spots as follows.
First, we derived rotational periods by the discrete Fourier transform of the whole light curves.
We here selected stars with their rotational periods more than 1 day and less than 30 day because too rapid or slow rotations complicate tracing local minima for a long time.
Investigating a dependence of rotational periods on lifetimes is not our main purpose in this paper, though it should be investigated in our future works.
Next, we obtained the smoothed light curve by using locally weighted polynomial regression fitting \citep[LOWESSFIT;][]{LOWESS} to remove flare signature and noise.
In the LOWESSFIT algorithm, a low degree polynomial is fitted to the data subset by using weighted least squares, where more weights are given to the nearby points. 
We used the $lowess$ function incorporated to the R package.
The fitting passbands were selected to be 4$\times P_{\rm rot}$ to avoid the over- and under-smoothing.
We detected the local minima as downward convex points of the smoothed light curve; i.e., the smoothed stellar fluxes $F(t)$ satisfy $F (t_{(\rm n-m)})$ $\rm <$ $F (t_{(\rm n-m-1)})$ and $F (t_{(\rm n+m)})$ $\rm <$ $F (t_{(\rm n+m+1)})$.
Here, m takes a value of [0, 1, 2], $t$ is time, and n is time step.
[A] We start to trace them from an arbitrary local minimum ($T_0$), and at first search another local minimum whose time is between $T_0+0.8\times P_{\rm rot}$ and $T_0+1.2\times P_{\rm rot}$ (see Figure \ref{fig:analysis} for the visual explanations of the procedure [A] $\sim$ [E]).
This range was determined to be able to cover the range of the solar-like differential rotation ($\Delta P/P_{\rm rot} \sim$ 0.2).
In case we find the next local minimum, we identify it as a next one ($T_1$).
If we successfully trace more than three local minima in the same manner, we identify them as a single spot candidate, and continue the tracking.
[B] After $T_2$, we decide to search the next local minimum as time in between $T_2+0.9\times P_{\rm spot}$ and $T_2+1.1\times P_{\rm spot}$, where $P_{\rm spot}$ is a rotational period of the spot candidate which is obtained in the procedure [A].
[C] In case that there are some observational data gaps, our algorithm is designed to be able to search a next local minimum until three rotational period ahead.
The algorithm also search the local minima before the start point ($T_0$) by the same manner.
If there is no local minimum in the next rotational phase, the algorithm stops to trace and switches to the next starting point ($T_0$).

After searching the spot candidates in a given star, the dubious candidates are automatically removed in the following procedure.
[D] First, for a given spot candidate, if there are other local minima within $\pm 0.35\times P_{\rm rot}$ of each $T_{\rm n}$ (n = 0$\sim$N, N+1 is the total number of local minima of the spot candidate), we remove the candidate because the spot area can be largely affected by the other spots.
[E] Second, we extrapolated the ``virtual" local minima for 3 period ahead and behind ($T_{-3,-2,-1}$ = $T_0$ - m$\times P_{\rm rot}$, $T_{N+1,N+2,N+3}$ = $T_N$ + m$\times P_{\rm rot}$, m is 1,2,3).
If there are other local minima within $ \pm 0.3\times P_{\rm rot}$ of each ``virtual" local minimum, we also remove the candidate because the spot can survive or combine with the other spot in longitude which is originally located at the different longitude.
The values of 0.35 and 0.3 are longitude separation over which we regard a pair of spots as located on reverse phase  (i.e. longitude separation $>$ 126$^{\circ}$, 108$^{\circ}$), and were determined based on the number of miss detections. If the values become smaller, the contaminations and mergings of other spots would not be negligible.
Lastly, we measure the local depth of the local minima from the near local maxima as an indicator of spot area.
%As we can expect from sunspot, the temporal variation of single star spots would show emergence and decay phase.
To investigate the nature of a single star spot as well as to remove the beating spot candidate, we also remove the spot candidates whose local depth variation do not show clear emergence and decay phases.
Here we simply use chi-squared test to judge whether it shows emergence or decay phase or not.

We applied the above automatical detection method to the 5356 solar-type target stars observed by the Kepler, and obtained 147 spot candidates.
Visually checking all the light curves, phase-time diagrams of local minima, and temporal variations of local depth, we selected spot candidates that satisfy the following conditions: 
(1) The temporal variations show clear emergence and decay phases. 
We use a threshold that the first spot area $A(T_0)$ and final one $A(T_{\rm N})$ should be smaller than 70 \% of the maximum size ($A_{\rm max}$), i.e., $A(T_0)$ $<$ 0.7$\times A_{\rm max}$ and $A(T_{\rm N})$ $<$ 0.7$\times A_{\rm max}$. 
See Section \ref{sec:ana:area} for the detailed definitions of the spot area $A(t)$.
This threshold was determined to exclude the doubtful spot candidates (e.g. a pair of beating spots).
For example, if we choose 90 \% as a threshold, we need to expect contaminations by considerable amount of dubious spot candidates.
If the light curve satisfies this condition, it is easy to accurately measure the emergence and decay rates and extrapolate the variations to estimate the lifetimes.
This threshold has been determined by trial and error.
(2) The temporal evolutions of spot area show apparent single peaks.
Note that we do not remove candidates whose temporal evolutions can be separated from the other spot candidates.
(3) The observational noise is apparently much smaller than the spot amplitude.
(4) The light curves show no apparent beat features during the lifetimes.
(5) The observational gaps do not largely disturb the detection of local minima.
(6) The local minima should not disappear during the lifetime for reasons other than observational gaps.
(7) The local minima do not include any apparent miss detections of local minima.
We conducted these treatments (1) -- (7) with manual checking by more than three of the authors.
According to this procedure, we successfully identified the temporal evolutions of 56 star spots listed in Table 1. 
As for the spot candidates, we also improved the light curve fittings.
To avoid overfitting the flare signals, we simply detected large flare-like signals by using the threshold of superflare detections based on \cite{2012Natur.485..478M}, and remove them with spline interpolation.
After this process, we fit the long-period spot modulations by LOWESSFIT, where the fitting parameter are manually adjusted for each light curve.
Then we re-detected the local minima and trace them in the same manner.
These re-detected values are listed in Table 1.

\begin{table*}\label{table:1}
\caption{Physical parameters of star spots and the host stars.}
\begin{center}
\scriptsize
%\scalebox{0.75}[0.75]{
%\scalebox{0.75}[0.75]{
\begin{tabular}{lcccccccccc}
\hline
No. & Kepler ID & $T_{\rm eff}$\commenta & log $g$\commentb & $R_{\rm star}$\commentc & $P_{\rm rot}$\commentd & $T$\commente & $A_{\rm max}$\commentf & $\Phi_{\rm max}$\commentf &  $d\Phi_{\rm e} / dt$\commentg  &  $d\Phi_{\rm d} / dt$\commentg \\
 &  & [K] &  & [$R_{\odot}$] & [day] & [day] & [$10^3$MSH] & [$10^{23}$Mx] & [$10^{20}$Mx/h] & [$10^{20}$Mx/h] \\
\hline
1 & $10186360^{}$ & 5994 &  4.48  &  0.92  &  7.3  & $ 73.7 \pm  8.3 $ & $ 2.7 _{- 0.6 }^{+ 0.9 }$ & $ 1.4 _{- 0.4 }^{+ 0.8 }$ & $ 1.0_{- 0.6}^{+ 1.0}$ & $ 1.4_{- 0.7}^{+ 1.3}$ \\
2 & $10328375^{1}$ & 5631 &  4.48  &  2.07  &  5.1  & $ 63.9 \pm  1.7 $ & $ 27.7 _{- 6.7 }^{+ 8.9 }$ & $ 72.1 _{- 27.1 }^{+ 43.6 }$ & $86.6_{-38.9}^{+67.4}$ & $102.2_{-43.4}^{+73.5}$ \\
3 & $10736868^{}$ & 5505 &  4.44  &  1.16  &  6.3  & $ 57.2 \pm  20.1 $ & $ 9.1 _{- 1.7 }^{+ 2.0 }$ & $ 7.5 _{- 2.2 }^{+ 2.9 }$ & $ 9.1_{- 4.9}^{+ 7.8}$ & $ 6.8_{- 2.6}^{+ 3.8}$ \\
4 & $10802309^{1}$ & 5403 &  4.61  &  1.43  &  2.0  & $ 65.0 \pm  21.7 $ & $ 17.5 _{- 3.5 }^{+ 4.1 }$ & $ 21.7 _{- 6.9 }^{+ 9.4 }$ & $26.8_{- 9.9}^{+14.4}$ & $15.3_{- 6.3}^{+ 9.5}$ \\
5 & $10818810^{}$ & 5869 &  4.38  &  0.96  &  6.0  & $ 17.0 \pm  5.1 $ & $ 3.6 _{- 0.6 }^{+ 0.9 }$ & $ 2.0 _{- 0.5 }^{+ 0.9 }$ & - & - \\
6 & $10936008^{}$ & 5271 &  4.58  &  0.75  &  9.3  & $ 37.2 \pm  18.5 $ & $ 2.2 _{- 0.7 }^{+ 1.0 }$ & $ 0.7 _{- 0.3 }^{+ 0.6 }$ & - & - \\
7 & $10969515^{}$ & 5380 &  4.50  &  0.95  &  5.5  & $ 58.6 \pm  10.5 $ & $ 4.9 _{- 0.9 }^{+ 1.2 }$ & $ 2.7 _{- 0.8 }^{+ 1.2 }$ & $ 5.5_{- 2.5}^{+ 4.1}$ & $ 1.9_{- 0.9}^{+ 1.5}$ \\
8 & $11033729^{}$ & 5399 &  4.59  &  0.74  &  10.2  & $ 244.4 \pm  19.3 $ & $ 4.8 _{- 0.9 }^{+ 1.1 }$ & $ 1.6 _{- 0.5 }^{+ 0.7 }$ & $ 0.3_{- 0.1}^{+ 0.2}$ & $ 1.1_{- 0.4}^{+ 0.6}$ \\
9 & $11046341^{}$ & 5597 &  4.39  &  0.84  &  16.3  & $ 111.1 \pm  29.9 $ & $ 6.3 _{- 1.1 }^{+ 1.3 }$ & $ 2.7 _{- 0.8 }^{+ 1.1 }$ & $ 1.6_{- 1.1}^{+ 1.8}$ & $ 1.7_{- 0.5}^{+ 0.7}$ \\
10 & $11080702^{}$ & 5307 &  4.58  &  0.88  &  9.2  & $ 86.7 \pm  3.7 $ & $ 4.2 _{- 0.5 }^{+ 0.6 }$ & $ 2.0 _{- 0.4 }^{+ 0.5 }$ & $ 2.3_{- 0.7}^{+ 1.0}$ & $ 1.4_{- 0.4}^{+ 0.6}$ \\
$\cdots$ & $\cdots$ & $\cdots$ & $\cdots$ & $\cdots$ & $\cdots$ & $\cdots$ & $\cdots$ & $\cdots$ & $\cdots$ & $\cdots$ \\
55 & $9408373^{}$ & 5782 &  4.49  &  0.91  &  5.1  & $ 31.7 \pm  1.6 $ & $ 2.8 _{- 0.5 }^{+ 0.7 }$ & $ 1.4 _{- 0.4 }^{+ 0.7 }$ & $ 2.9_{- 1.4}^{+ 2.4}$ & $ 5.0_{- 2.4}^{+ 4.2}$ \\
56 & $9579266^{}$ & 5164 &  4.60  &  0.83  &  10.6  & $ 63.2 \pm  10.0 $ & $ 2.6 _{- 0.7 }^{+ 1.0 }$ & $ 1.1 _{- 0.4 }^{+ 0.7 }$ & - & $ 0.8_{- 0.4}^{+ 0.7}$ \\
\hline
\multicolumn{11}{l}{$^1$ Subgiant or main-sequence binary candidates \citep{2018ApJ...866...99B}. $^2$ cool main-sequence binary candidates \citep{2018ApJ...866...99B}. }\\
\multicolumn{11}{l}{\commenta Stellar effective temperature taken from Kepler Input Catalog \citep[Kepler Data Release 25 Notes,][]{2016KeplerDR25}. }\\
\multicolumn{11}{l}{\commentb Stellar surface gravity taken from Kepler Input Catalog. \commentc Corrected stellar radii by \cite{2018ApJ...866...99B}.} \\
\multicolumn{11}{l}{\commentd Stellar rotational periods. \commente Lifetimes of star spots. \commentf Maximum star spot area and magnetic flux in the unit of MSH and Mx, respectivitly.} \\
\multicolumn{11}{l}{\commentg Emergence and decay rates of star spots in the unit of Mx per hour.}
\end{tabular}
%}
\end{center}
\end{table*}

\subsection{Area Estimation}\label{sec:ana:area}
We estimated the area of star spots based on the brightness depth of each local minimum ($\Delta F$) from the nearby local maximum.
Deriving the spot area from the $\Delta F$ requires measurements of the spot temperature \citep[e.g.][]{1985ApJ...289..644P}.
However, since the Kepler conducted single-bandpass observations, we cannot distinguish a decrease of the spot temperature from an increase of the spot area and vice versa.
Here, we used the following empirical relation of spot temperature as a function of stellar effective temperature.
According to \cite{2017PASJ...69...41M}, the area ($A_{\rm spot}$) can be derived as a function of the normalized amplitude ($\Delta F/F_{\rm star}$), stellar effective temperature ($T_{\rm star}$), and stellar radius ($R_{\rm star}$):
\begin{eqnarray}
A_{\rm spot} &=& \left( \frac{R_{\rm star}}{R_{\odot}} \right)^2\frac{T_{\rm star}^4}{T_{\rm star}^4-\{T_{\rm star}-\Delta T(T_{\rm star})\}^4}\frac{\Delta F}{F_{\rm star}},\\
\Delta T (T_{\rm star}) &=& T_{\rm star}-T_{\rm spot} = 3.58 \times 10^{-5} T_{\rm star}^2 + 0.249 T_{\rm star}-808,
\end{eqnarray}
where $\Delta T$ is temperature difference between photosphere and spot derived based on \cite{2005LRSP....2....8B}.
The spot temperature is basically estimated by the Doppler imaging technique of several main-sequence stars.
Since this relation is just an empirical one, the spot area can change if the actual spot temperature varies.
However, the variation of the temperature by ±500 K (±1000 K) could vary the spot area by only 11 \% (23 \%).
Therefore, our results would not be significantly affected by the assumption of temperature.
Here, the stellar effective temperature (T) is based on the Kepler Input Catalog \citep[\textit{Kepler} Data Release 25,][]{2016KeplerDR25}.
As for the stellar radius, we use the radius values updated by using recent \textit{Gaia} satellite Data Release 2 \citep{2018ApJ...866...99B,2018A&A...616A...2L}.

Note that the estimated area can be somewhat underestimated due to inclinations of the stellar rotational axes and the contaminations of brightness from other spots.
The latter may be corrected by modeling the light curves, but we simply use the local depth of light curves as an indicator of spot area.
Moreover, the faculae on the stellar surface can also contribute to the over- and under-estimation of the star spot area.
In Section \ref{sec:md}, the uncertainties by those effects are clarified, while they are not incorporated to the estimations in this paper.
In Appendix \ref{app1}, we simply evaluated accuracies of the area estimations on the basis of the Sun-as-a-star analysis.
%This is beyond the scope of this work, but would be addressed in the future work.

\subsection{Lifetime Estimation}
We measured the lifetimes ($T_{1}$) of the 56 star spots based on how long the local minima are detectable (i.e., $t_N-t_1$).
The lifetimes ($T_{1}$) can be, however, underestimated because the detectable limits of amplitude can largely suffer from noises and contaminations of other spots.
%Also, the beginnings and ends of the local minima might be not clear as for some star spots due to the contamination of the other spots.
Therefore, we fitted the emergence and decay phases with linear relations (solid lines in Figure \ref{fig:oc0}(d)) and estimated the lifetimes ($T_{2}$) from spot emergence to disappearance.
%We have corrected the lifetimes for only spots whose emergence and decay phases have more than 3 points and significant positive and negative slope, respectively.
%In case $T_1>T_2$, we determined $T_2 = T_1$.
In the following section, we defined the lifetime ($T$) as ($T_1+T_2$)/2, the lower limit as $T_1$, and the upper limit as $T_2$.
%When the spots do not show clear emergence or decay phase, we do not extrapolate the lifetimes.
Note that $T_2$ are not exactly the upper limit values, but extrapolated ones by assuming the linear emergence and decay.
As mentioned in the following, it may be better to fit by assuming parabolic decay.
However, the decay phases do not necessarily show the clear parabolic decay curves, so we use this assumption.

\subsection{Calculation of Emergence and Decay Rates of Star Spots}
We estimated the emergence ($d\Phi_{\rm e}/dt$) and decay rates ($d\Phi_{\rm d}/dt$) of the star spots based on the variation rates of star spot area.
The variation rates are considered to be better indices when comparing with sunspot properties because they are unaffected by the detectable limits of local minima unlike the lifetimes.
We derived the variation rates of star spot area (emerging rate $dA_{\rm e}/dt$ and decay rate $dA_{\rm d}/dt$) by applying the linear regression fitting method to spot area variations.
Many papers have reported variation rates of sunspots in the unit of Mx = $\rm G\cdot cm^2$ \citep[e.g.,][]{2017ApJ...842....3N}.
Taking this fact into account, we calculated the emergence and decay rates by assuming that the mean magnetic field of star spots is 2000 G considering the typical magnetic field strength of sunspots \citep{2003A&ARv..11..153S}:
\begin{eqnarray}
\frac{d \Phi_{\rm e,d}}{dt} &=& 2000\times \frac{dA_{\rm e,d}}{dt}. 
\end{eqnarray}
%This assumption can lead to some errors, but enables us to simply compare with sunspots.
The error values are mainly estimated based on the errors of the fitted slopes, stellar effective temperature, and stellar radius.

\section{Result} \label{sec:res}

\subsection{Temporal Evolutions of Star Spots}
Figure \ref{fig:oc0} shows an example of the temporal evolutions of star spots in our catalog.
Figure \ref{fig:oc0}(a) shows a light curve and circles are the detected local minima.
Figure \ref{fig:oc0}(c) shows time-phase diagram, where the vertical axis is the rotational phase of the detected local minima.
The gray solid line is an individual spot component that we detected.
Figure \ref{fig:oc0}(d) shows the temporal evolution of the local depth of each local minima that correspond to the star spot area.
As described in Section \ref{sec:ana}, we selected star spots showing clear emergence and decay, and such features can be also seen in Figure \ref{fig:oc0}.

As a result of these analyses, the area of the detected star spots are 1500--23000 MSH at the maximum, and the lifetimes are 10--350 days (see Table 1).
The lifetime of one year is the longest ever observed for the solar-type stars \citep[$\lesssim$ 200 days in the previous studies:][]{2014ApJ...795...79B,2015PhDT.......177D,2017MNRAS.472.1618G}.
In the case of the sunspots, there is a notable asymmetry in emergence and decay phase as mentioned in Section \ref{sec:int}.
However, in the case of star spots, the averaged emergence rates $6.6\times10^{20}$ $\rm Mx\cdot hr^{-1}$ ($11$ $\rm MSH\cdot hr^{-1}$) are not so much different from the averaged decay rates $5.2\times10^{20}$ $\rm Mx\cdot hr^{-1}$ ($8.5$ $\rm MSH\cdot hr^{-1}$).
Interestingly, in several spots, emergence phase are longer than the decay phase (see the figures in the online version of the Journal for the detail).
%The emergence phases surprisingly take several rotation periods, which are much longer than sunspots \citep[typically less than 5 days;][]{1993SoPh..148...85H}. 
There is a possibility that not only the area but also lifetimes can have some uncertainties caused by the data sensitivity and analytical method.
The uncertainties are discussed in Section \ref{sec:md}.

\subsection{Star Spot Properties versus Stellar Properties}

\begin{figure}[htbp]
\begin{center}
\includegraphics[scale=0.7]{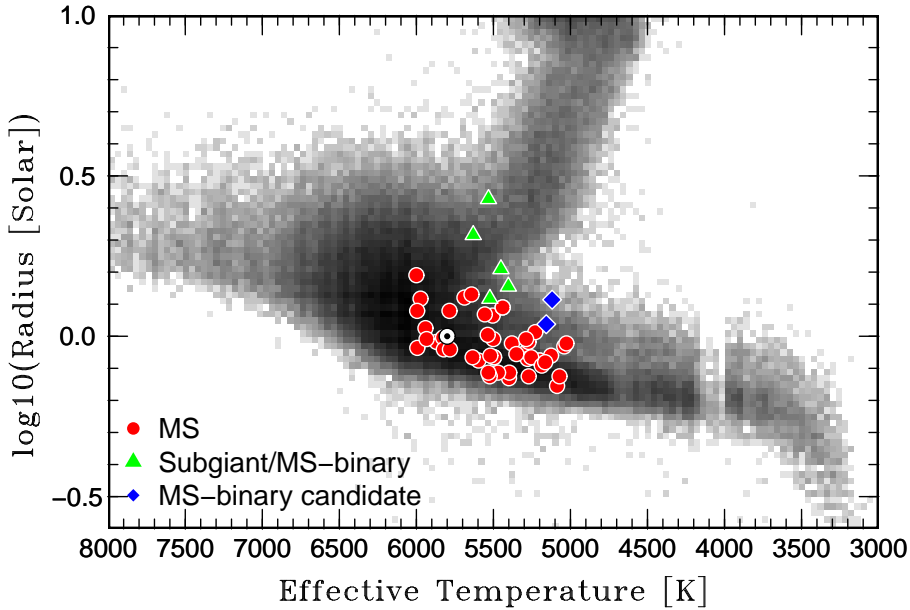}
\caption{Radius versus effective temperature.
Background black shades indicate distributions of \textit{Kepler} stars with revised radii based on Gaia DR2 parallaxes presented in \cite{2018ApJ...866...99B}.
Red circles, green triangles, and blue diamonds are solar-type stars in our catalog, which had been judged as solar-type stars in Kepler Input Catalog \citep[\textit{Kepler} Data Release 25 Notes,][]{2016KeplerDR25}.
They are classified as main-sequence stars, subgiant stars/main-sequence binary stars, and main-sequence binary stars, respectively \citep{2018ApJ...866...99B}. Solar value is also plotted for reference (the circled dot).}
\label{fig:hrdiagram}
\end{center}
\end{figure}

Figure \ref{fig:hrdiagram} shows stellar radius as a function of the effective temperature (i.e., HR diagram).
Background black shades indicate distributions of \textit{Kepler} stars and colored symbols indicate our catalog.
The vertical axis values are plotted with revised radii based on quite recent Gaia DR2 parallaxes provided by \cite{2018ApJ...866...99B}.
According to \cite{2018ApJ...866...99B}, four of our ``solar-type" stars are classified into subgiant stars or main-sequence binaries stars (green triangles), and two are classified into main-sequence binary star candidates (blue diamonds), although they are solar-type stars according to the Kepler Input Catalog.
In the following figures and discussions, we only focus on the temporal evolutions of star spots on main-sequence stars.

\begin{figure}[htbp]
\begin{center}
\includegraphics[scale=0.7]{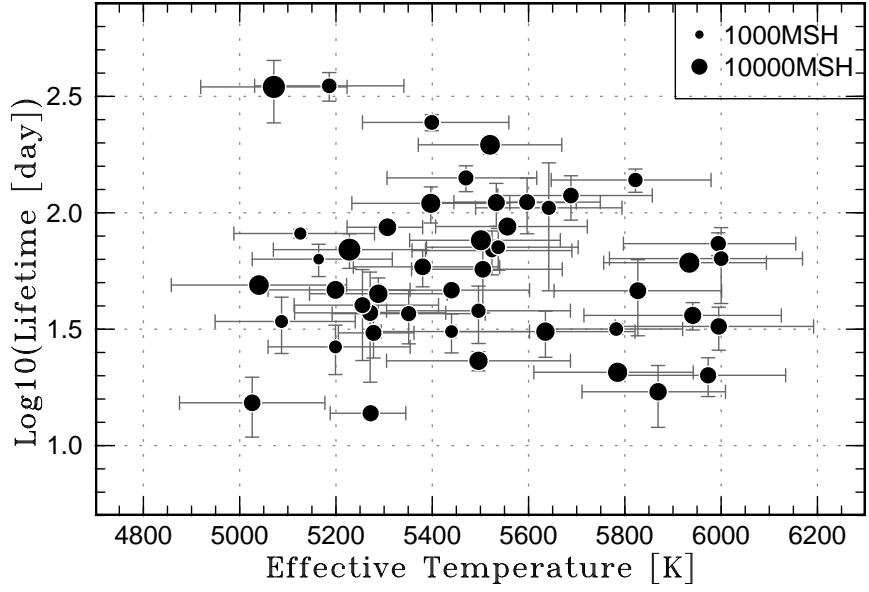}
\caption{Comparison between the stellar effective temperatures ($T_{\rm eff}$) and lifetimes of star spots.
The size of each circle corresponds to the maximum area of the star spot.
The temperatures and the error bars are taken from Kepler Input Catalog \citep[\textit{Kepler} Data Release 25 Notes,][]{2016KeplerDR25}.}
%We have not plotted stars which have been classified to non main-sequence stars by \cite{2018arXiv180500231B}.}
\label{fig:temp}
\end{center}
\end{figure}

Figure \ref{fig:temp} shows a comparison between stellar effective temperature and lifetimes of star spots.
There seems to be no clear temperature dependence even for a given spot size, although \cite{2017MNRAS.472.1618G} have reported that cooler stars have spots that last much longer.
This might be due to our small number of samples and small range of temperature. We focus only on G-stars while they analyzed F, G, K, M-stars.
%The temperature dependence will be discussed in the next paper.
In this paper, we do not discuss the relation between stellar effective temperature and star spot properties because of shortage of samples and range of stellar properties.
Nevertheless, the dependence of temperature on the lifetimes is quite interesting to investigate the role of differential rotation and convection in the spot fragmentations.
This dependence is subjected to consideration in the future study.

\begin{figure}[htbp]
\begin{center}
\includegraphics[scale=0.7]{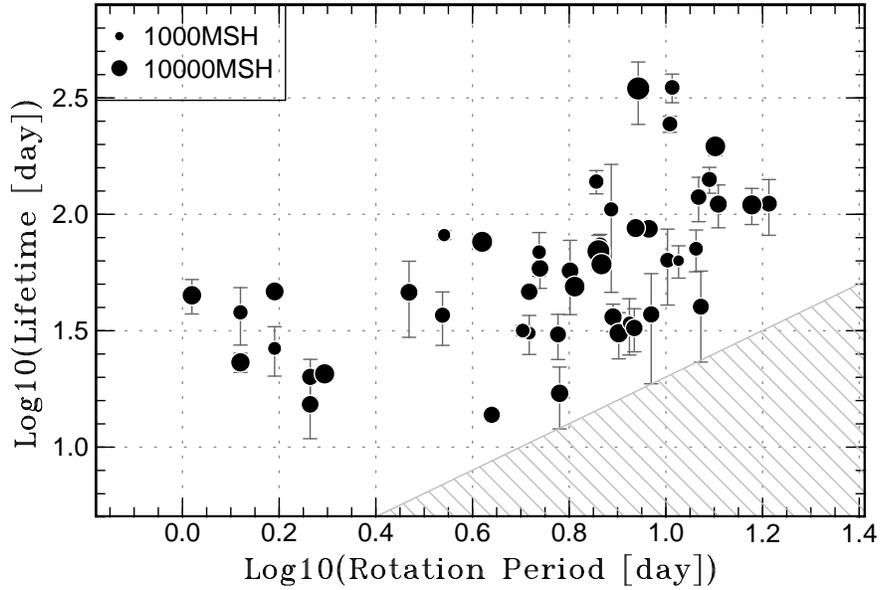}
\caption{Comparison between the stellar rotation period ($P$) and lifetimes of star spots.
The size of each circle represents the maximum area of the star spot.}
%We have not plotted stars which have been classified to non main-sequence stars by \cite{2018arXiv180500231B}.}
\label{fig:rotperiod}
\end{center}
\end{figure}

Figure \ref{fig:rotperiod} shows a comparison between rotational periods and lifetimes of star spots and there looks to have a positive correlation.
Please note that there are an undetectable region because our algorithm can detect spots whose lifetimes are longer than a couple of rotational periods.

%Please note, however, that this may be attributed to our analytical biases.
%In the case of slowly rotating stars, the time cadences of the detectable local minima are worse, so that the measurable lifetimes should be longer than those in the case of rapidly rotating ones (see undetectable region in Figure \ref{fig:rotperiod}).
%Besides, rapidly rotating stars are indicated to have strong surface differential rotations compared to slowly rotating ones \citep[e.g.,][]{2011ApJ...740...12H,2016MNRAS.461..497B}.
%This makes it difficult to trace the local minima for a long time.
%As a result, the detectable spots can be limited to only short-lived spots in the case of rapidly rotating stars.
%These analytical biases can contribute to the apparent positive correlation.

\subsection{Comparison with Sunspot I: Lifetime versus Maximum Area}\label{sec:res:arealife}

\begin{figure}[htbp]
\begin{center}
\includegraphics[scale=0.7]{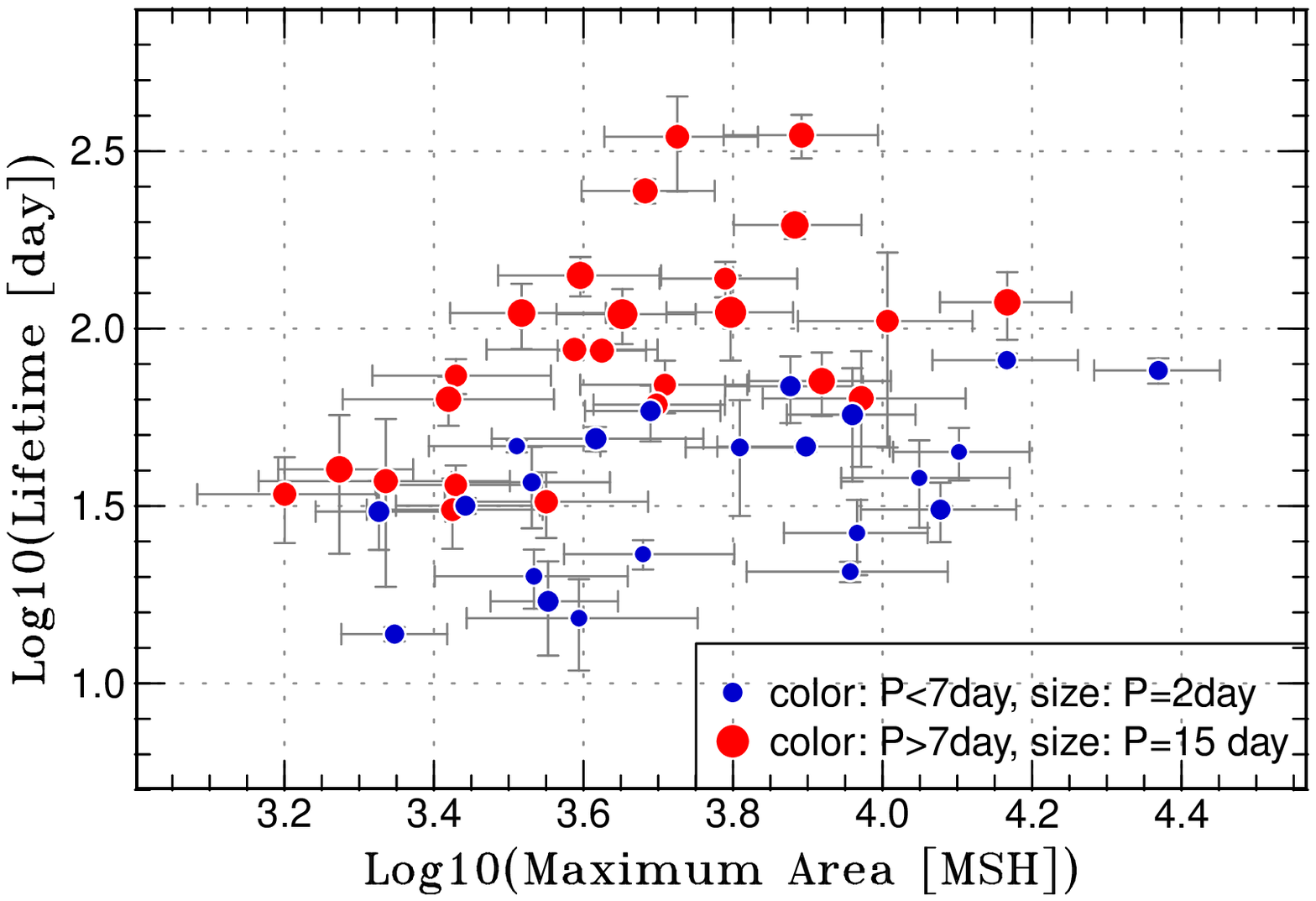}
\caption{Comparison between maximum areas and lifetimes ($T$) of the detected star spots on solar-type stars. 
%Circles correspond to main-sequence stars based on classification by \cite{2018arXiv180500231B}. 
%The upper values of lifetimes are extrapolated lifetimes ($T_2$), lower values are measured one ($T_1$), and the symbols correspond to the average ($T=(T_1+T_2)/2$).
The area is plotted in the unit of million solar hemisphere (1MSH = $10^{-6} \times 2\pi R_{\odot}^2$).
The size of each circle represents the stellar rotational period.
Blue and red colors correspond to the spots with $P_{\rm rot}<$ 7 day and $P_{\rm rot}>$ 7 day, respectively.}
\label{fig:lifearea:star}
\end{center}
\end{figure}

Figure \ref{fig:lifearea:star} shows a comparison between the maximum area and lifetimes of the star spots in our catalog.
We plotted the star spot data for slowly rotating stars ($P_{\rm rot}$ $>$ 7 days) and rapidly rotating ones ($P_{\rm rot}$ $<$ 7 days), separately.
There seems to be a weak positive correlation between them.

\begin{figure}[htbp]
\begin{center}
\includegraphics[scale=0.7]{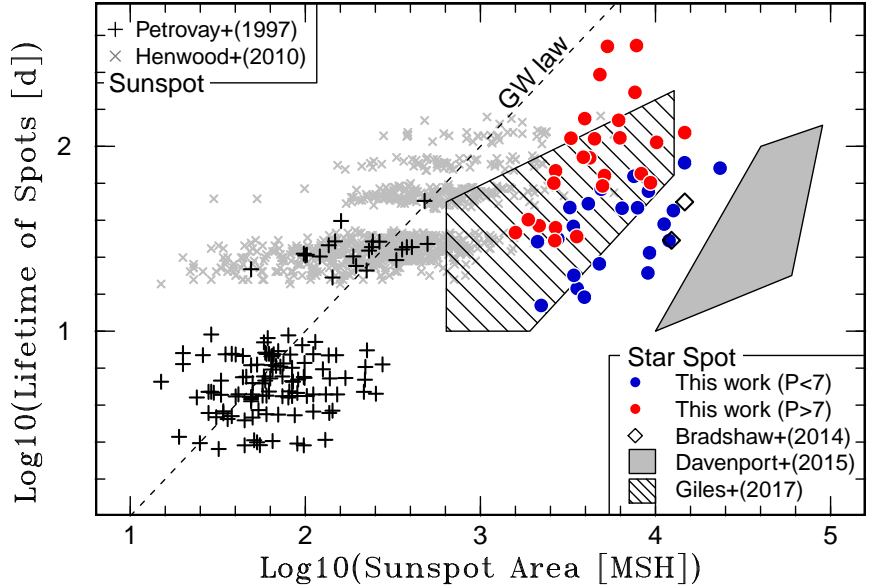}
\caption{Comparison between maximum spot area and lifetime of sunspots and star spots on solar-type stars.
Black and gray crosses are sunspots data taken from \cite{1997SoPh..176..249P}  and \cite{2010SoPh..262..299H}, respectively.
The dashed line indicates the solar GW relation ($A = DT$, $D\sim 10$ $\rm MSH\cdot day^{-1}$).
%Note that, as for the data of \cite{2010SoPh..262..299H}, we have excluded the sunspots whose temporal evolutions show multiple growths.
Blue and red circles correspond to the spots analyzed in this study with $P_{\rm rot}<$ 7 day and $P_{\rm rot}>$ 7 day, respectively.
Open diamonds are star spots on G-type stars (Kepler-17 and CoRoT-2) taken from \cite{2014ApJ...795...79B}, which were estimated by using exoplanet transits.
A grayed filled region indicates star spots on Kepler-17 detected in \cite{2015PhDT.......177D}.
Note that the area and lifetimes of star spots on Kepler-17 are also estimated in \cite{2014ApJ...795...79B}, but there is a factor gap in spot area between \cite{2014ApJ...795...79B} and \cite{2015PhDT.......177D}.
A region filled with diagonal lines indicates the result of \cite{2017MNRAS.472.1618G}.}
\label{fig:lifearea}
\end{center}
\end{figure}

Figure \ref{fig:lifearea} shows a comparison between the maximum area and lifetime of sunspots and star spots on solar-type stars.
Black and gray points are sunspot data taken from \cite{1997SoPh..176..249P} and \cite{2010SoPh..262..299H}, respectively.
These sunspot data are basically measured by using the Debrecen Photoheliographic Results and Greenwich Photoheliographic Results, respectively.
They are available upon the databases recording day--to--day individual sunspot areas.
Note that, since the identification of recurrent sunspots is based only on its longitude and latitude, the succeeding spot emergence in the decaying active regions can be identified as a single spot.
For example, \cite{1984SoPh...93..181K} have reported long-lived sunspot groups surviving during 8 solar rotation by use of Greenwich Photoheliographic Results.
The temporal development of spot area shows, however, several peaks, which indicates successive episodes of spot emergence in the same region.
Our main purpose is to reveal a star spot physics from the basic sunspot physics, and the comparison with such sunspots with several emergence would lead to more complex discussions.
Therefore, as for the data of \cite{2010SoPh..262..299H}, we have excluded the sunspots whose temporal evolutions show multiple growths for matching our star spots that have only simple emergence-decay patterns.
The resulting lifetimes of sunspots are up to 6 solar rotational period ($\sim$200 days) and the area is up to $\sim$6000 MSH.
The dashed line indicates the GW relation mentioned in Section \ref{sec:int} ($A = DT$, $D\sim 10$ $\rm MSH\cdot day^{-1}$).

In Figure \ref{fig:lifearea}, we also plotted star spots on solar-type stars with red and blue filled circles for slowly ($P_{\rm rot}$ $>$ 7 days) and rapidly rotating stars ($P_{\rm rot}$ $<$ 7 days), respectively. % only for stars classified to main-sequence stars by \textit{Gaia} DR2.
We found that the lifetimes of large star spots (10$\sim$350 day) are shorter than those expected from the GW relation (300$\sim$1000 day).
This trend is similar to the results reported in the other previous researches.
We plotted the star spots that were reported in \cite{2014ApJ...795...79B}, \cite{2015PhDT.......177D}, and \cite{2017MNRAS.472.1618G}.
%\cite{2014ApJ...795...79B} roughly estimated the star spot lifetimes based on the brightness variation during exoplanet transit in Kepler-17 and COROT-2.
%\cite{2015PhDT.......177D} also analyze the transit light curve of Kepler-17, and showed a good correlation between the lifetime and spot area.
%Note that the difference between the result of the above two authors may mean that the area estimation can be dependent on the analytical method.
%Besides, \cite{2017MNRAS.472.1618G} derived the decay rate of auto-correlation function as a good indicator of star spot lifetime for huge amount of Kepler stars.
%They showed that there is a good correlation between typical decay timescale and typical brightness variation amplitude.
Note that  the data of \cite{2017MNRAS.472.1618G} are given in the unit of brightness variation amplitude in their paper, so we plotted their G-type star data by assuming that all of the effective temperatures and radii are the same as the solar values.
As a result, we found that our results are consistent with that of \cite{2017MNRAS.472.1618G}, and partly similar to that of \cite{2014ApJ...795...79B}.
The spot area of \cite{2015PhDT.......177D} are much larger than our data, while the lifetimes are similar to our results.

%Including our results, the lifetime of star spot show short lifetime than those expected from sunspots data.
%These analyses can provide detailed information on the mechanism of temporal evolutions of star spots.
%However, it is notable that these can include some observational or analytical problem.
%For example, star spot would become difficult to detect as the spot become small, which lead to underestimation of the lifetimes.
%In order to overcome this, we extrapolated the lifetime by assuming linear emergence and decay, but there is no theoretical support to this assumption.
%Moreover, there is a possibility that there is a selection biases.
%The long-lived spot with lifetime with $\sim$1,000 days can be missed because the Kepler observational period is limited to only 4 years ($\sim$1,500 days).
%For these reasons, the straightforward comparison is necessary to take care.

\subsection{Comparison with Sunspot II: Emergence Rates versus Maximum Flux}
\begin{figure}[htbp]
\begin{center}
\includegraphics[scale=0.7]{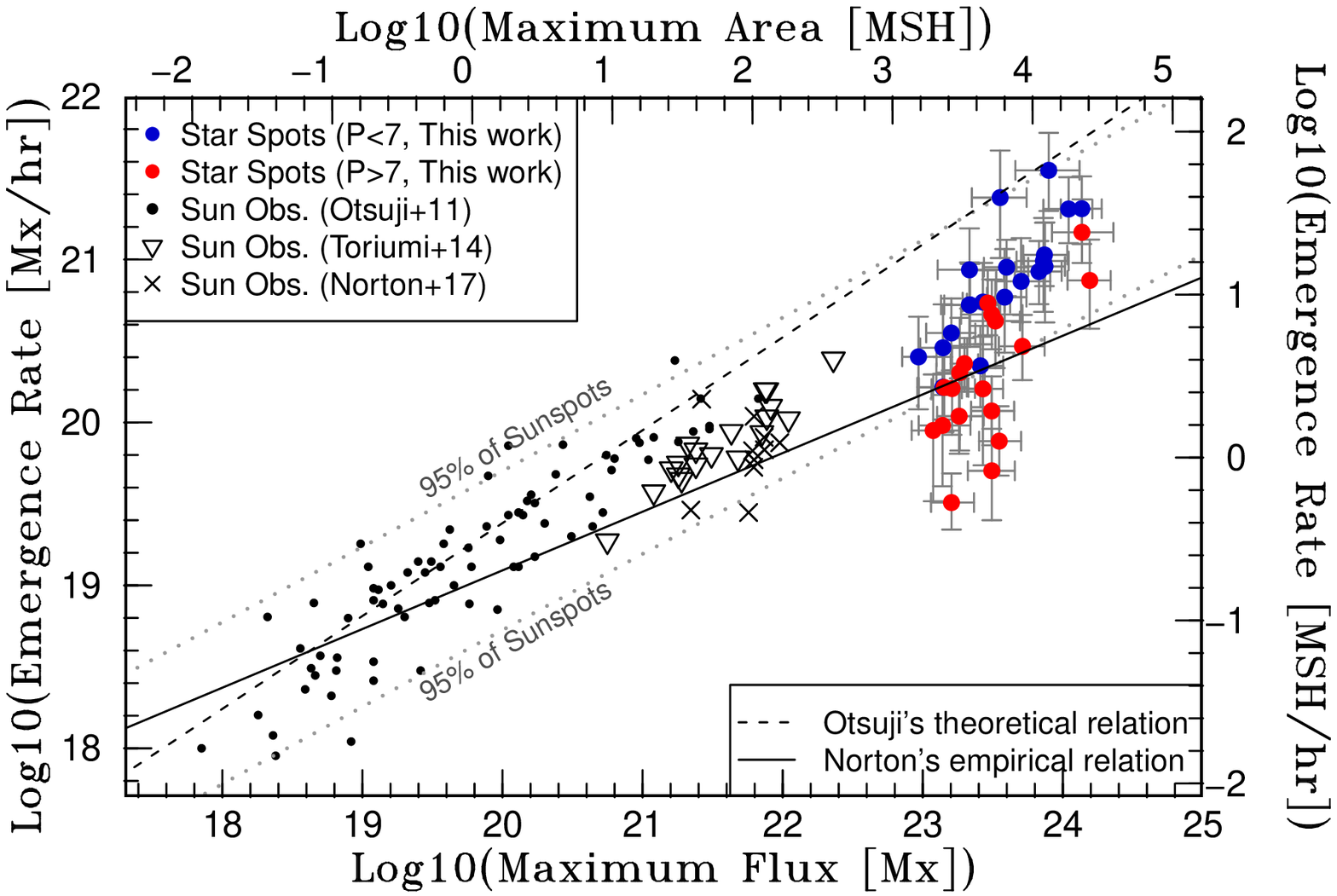}
\caption{Comparison between maximum magnetic flux ($\Phi$) and emergence rate ($d\Phi_{\rm e}/dt$) of sunspots and star spots on solar-type stars.
Black points, downward triangles, and crosses are sunspots observational data taken from \cite{2011PASJ...63.1047O}, \cite{2014ApJ...794...19T}, and \cite{2017ApJ...842....3N}, respectively.
%Red filled circles are star spots on solar-type stars measured in this study.
Blue and red circles correspond to the spots analyzed in this study with $P_{\rm rot}<$ 7 day and $P_{\rm rot}>$ 7 day, respectively.
%Blue asterisks are the result of numerical simulations \citep{2008ApJ...687.1373C,2014ApJ...785...90R,2017ApJ...838..113L}.
A solid ($d\Phi_{\rm e}/dt \propto \Phi^{0.3}$) and dashed line ($d\Phi_{\rm e}/dt \propto \Phi^{0.5}$) is a scaling relation derived by \cite{2017ApJ...842....3N} and \cite{2011PASJ...63.1047O}, respectively. 
The standard deviations of the star spot data compared with the scaling laws are 7.1$\times10^{20}$ and 9.5$\times10^{20}$ for Norton's and Otsuji's scaling law, respectively. 
Dotted lines are upper and lower of the 95\% confidence interval for the sunspot observations. 
76\% of the star spots are located in the extrapolation of this 95\% confidence interval.}
\label{fig:emerge}
\end{center}
\end{figure}

Figure \ref{fig:emerge} shows a comparison between the emergence rates and maximum fluxes of sunspots and star spots on solar-type stars.
%The vertical axis represents the emergence rates in the unit of Mx per hours.
The solar values are based on the magnetogram observation carried out by  \cite{2011PASJ...63.1047O}, \cite{2014ApJ...794...19T}, and \cite{2017ApJ...842....3N}.
%\f{purple}{Blue points are the result of recent numerical simulations performed by \cite{2008ApJ...687.1373C}, \cite{2014ApJ...785...90R}, and \cite{2017ApJ...838..113L}}.
%Note that we empirically assumed that the mean magnetic field strength of star spots is 2000 G.
The grayed dotted lines are the 95\% confidence intervals of the sunspot observational data.
As a result of comparison, 76\% of the star spots are consistently included inside the extrapolated 95\% confidence intervals of sunspots.
%, implying that the emergence rates of star spots are in most cases consistent with those of sunspots.
%The emergence rate of star spots are located on the middle between an empirical line ($d\Phi_{\rm e}/dt \propto \Phi^{0.3}$) derived by \cite{2017ApJ...842....3N} and a simple theoretical line ($d\Phi_{\rm e}/dt \propto \Phi^{0.5}$) derived by \cite{2011PASJ...63.1047O}.
Particularly, the emergence rates of star spots are consistent with an empirical line ($d\Phi_{\rm e}/dt \propto \Phi^{0.3}$) derived by \cite{2017ApJ...842....3N}.
The standard deviations of difference between star spot observations and the empirical line are 7.1$\times 10^{20}$ $\rm Mx\cdot h^{-1}$.
On the other hand, they are mostly smaller than those expected from a simple theoretical line ($d\Phi_{\rm e}/dt \propto \Phi^{0.5}$) derived by \cite{2011PASJ...63.1047O}.
%The details of the theoretical scaling law will be discussed in Section \ref{sec:dis:emerge}.
The standard deviations of difference between star spot observations and the theoretical line are 9.5$\times 10^{20}$ $\rm Mx\cdot h^{-1}$.
%This means that the star spot observational data are more consistent with the empirical relation $d\Phi_{\rm e}/dt \propto \Phi^{0.3}$ \citep{2017ApJ...842....3N}.
Moreover, the emergence rates of star spots looks to be dependent of the stellar rotational period dependence. The star spots on slowly rotating stars shows relatively small values, which is similar to the case of the lifetimes of star spots.
%As mentioned in Section \ref{sec:res:arealife}, this trend can be due to the analytical biases, so we will not mainly discuss the detailed distributions but focus on the overall statistical properties.
%The rotational periods of all stars showing low emergence rates are found to be relatively slowly rotating stars $P_{\rm rot}$ $>$ 7 days, which can result in this steep trend can be due to the rotational dependence.
%This implies that the feature can be an apparent trend.
%Here, we will not mainly discuss the detailed slope but focus on the overall statistical values compared with sunspot distributions in this paper.

\subsection{Comparison with Sunspot III: Decay Rate versus Maximum Flux}
\begin{figure}[htbp]
\begin{center}
\includegraphics[scale=0.7]{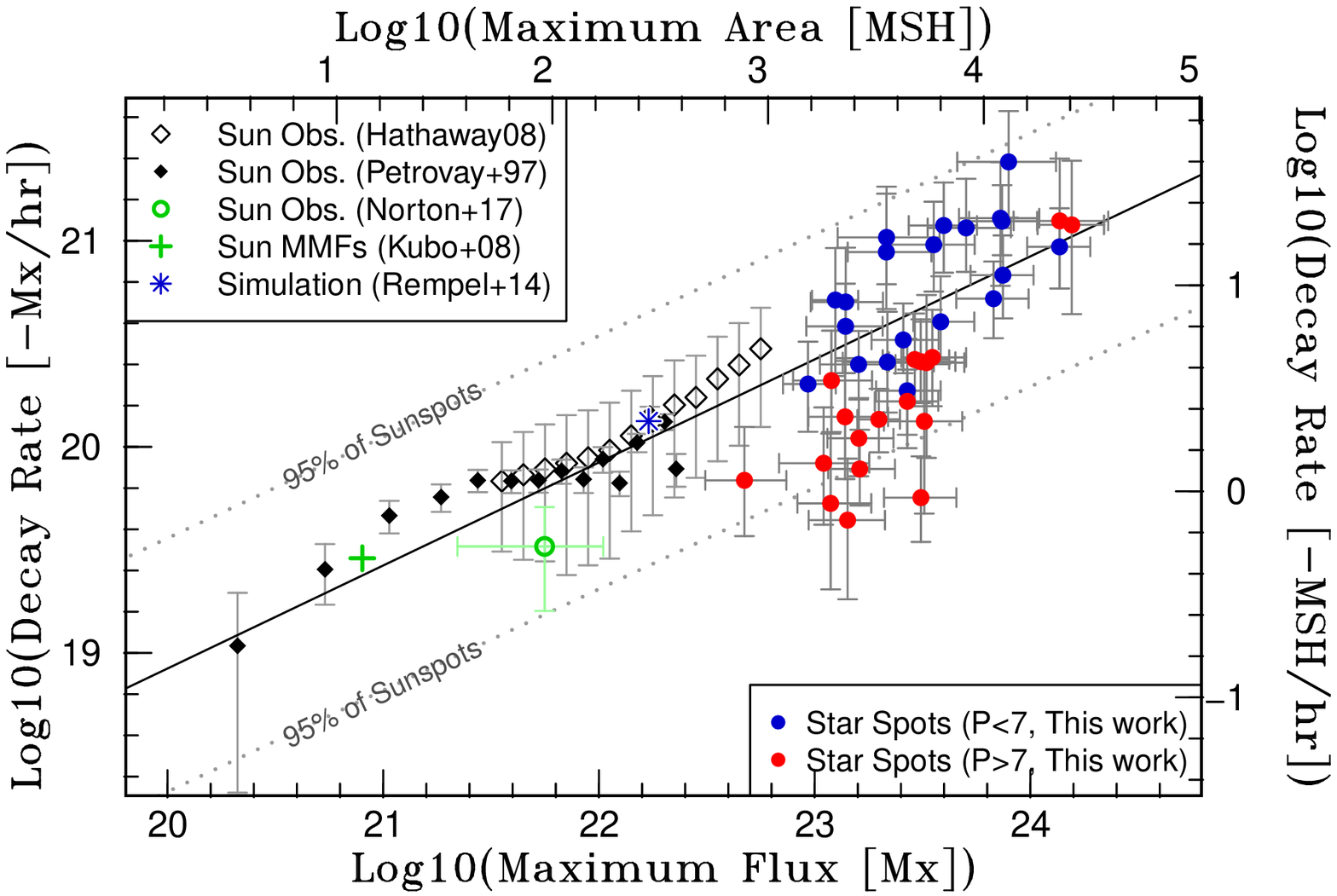}
\caption{Comparison between maximum magnetic flux ($\Phi$) and decay rate ($d\Phi_{\rm d}/dt$) of sunspots and star spots on solar-type stars.
Black open and filled diamonds are sunspot's observations by \cite{2008SoPh..250..269H}  and \cite{1997SoPh..176..249P}, respectively.
Note that the error bars of \cite{2008SoPh..250..269H} indicate 1-$\sigma$ level of their data, and those of \cite{1997SoPh..176..249P} indicate 3-$\sigma$ level.
As for the data of \cite{1997SoPh..176..249P}, the area is not the maximum area but the area when the decay rates are measured.
A green circle shows sunspot observations with \textit{SDO} magnetogram \citep{2017ApJ...842....3N}, and a green cross is a sunspot decay rate owing to moving magnetic features \citep[MMFs;][]{2008ApJ...686.1447K}.
%Red filled circles are star spots on solar-type stars measured in this study.
Blue and red circles correspond to the spots analyzed in this study with $P_{\rm rot}<$ 7 day and $P_{\rm rot}>$ 7 day, respectively.
Blue asterisks are the result of a simulation performed by \cite{2014ApJ...785...90R}.
Dotted lines are upper and lower of 95\% confidence levels for the sunspot observations \citep{2008SoPh..250..269H}.
89\% of the star spots are located in this extrapolation of 95\% confidence interval.
A solid line is the line of $d\Phi_{\rm d}/dt \propto \Phi^{0.5}$, where the absolute values are derived based on mean values of the sunspot observations \citep{2008SoPh..250..269H}.}
\label{fig:decay}
\end{center}
\end{figure}

Figure \ref{fig:decay} shows a comparison between the decay rates and maximum fluxes of sunspots and star spots on solar-type stars.
Black points are based on the visible sunspot observations \citep{2008SoPh..250..269H,1997SoPh..176..249P}, while the green points are based on the magnetogram observations \citep{2008ApJ...686.1447K,2017ApJ...842....3N}.
%Blue points are the result of simulations performed by \cite{2014ApJ...785...90R}.
The grayed dotted lines are the 95\% confidence intervals of the sunspot observational data of \cite{2008SoPh..250..269H}.
The order of the confidence intervals is about one order of magnitude, which is roughly the same as those of emergence rates.
As a result, the decay rates of star spots are also consistent with those of sunspots, while some parts are smaller than those expected from the sunspot distributions.
89\% of the star spots are included inside the extrapolated 95\% confidence intervals of sunspots data of \cite{2008SoPh..250..269H}.
The decay rates, including simulations, are roughly on the same lines (a solid line in Figure \ref{fig:decay}; $d\Phi_{\rm d}/dt \propto \Phi^{0.5}$) over a wide range of magnetic flux ($10^{21}$--$10^{24}$ Mx, 20--20000 MSH).
%The emergence rates of star spots are roughly consistent with or smaller than expected from sunspot data.
Also, a rotational period dependence can be also seen in the case of decay rates.

\section{Discussion} \label{sec:dis}
%%Emergence%%
\subsection{Emergence of Star Spots}\label{sec:dis:emerge}
%In many cases, lifetimes has been mainly discussed in terms of temporal evolutions of star spots.
%On the other hands, the emergence and decay rates of star spots can be considered to be more meaningful than the lifetimes because they are not affected by the analytical problem mentioned above.
We showed that the emergence and decay rates of star spots as a function of maximum fluxes are mostly consistent with those extrapolated from the sunspot observations.
This may suggest that the temporal evolutions of sunspot and star spot can be universally explained by the same underlying physical processes.
As for the emergence, we found that the emergence rates of star spots favor the empirical relation $d\Phi_{\rm e}/dt \propto \Phi^{0.3}$ \citep{2017ApJ...842....3N}, but are smaller than those expected from the theoretical scaling relation $d\Phi_{\rm e}/dt \propto \Phi^{0.5}$ \citep{2011PASJ...63.1047O}.

%have derived this empirical relation by fitting the result of several researches, but they did not provide any theoretical understandings of the scaling law.
\cite{2011PASJ...63.1047O} derived the simple theoretical scaling law $d\Phi_{\rm e}/dt \propto \Phi^{0.5}$ by assuming that (1) the emerging flux is self-similar in its size ($\Phi_{\rm e}\propto wh\propto w^2$: $w$ is horizontal, and $h$ is vertical length of its cross section) and (2) the rise velocity ($v$) is independent of its size ($d\Phi_{\rm e}/dt \propto vw \propto w$).
In spite of these rough assumptions, the scaling law agrees with their own observational data.

If we discuss the results on the basis of the theoretical scaling law ($d\Phi_{\rm e}/dt \propto \Phi^{0.5}$), there is a possibility that the emergence rates of star spots are suppressed to lower values for some reasons.
One explanation for the lower values is that the observed large star spots ($\sim 3\times 10^{23}$ Mx, $\sim 5000$ MSH) can be conglomerates of relatively smaller sunspot-scale spots (e.g., $\sim 10^{22}$ Mx, $\sim 160$ MSH). 
If the large star spots are conglomerates of smaller spots and the smaller spots emerge successively, the smaller emergence rates can be understood by extrapolating the emergence rates of sunspots.

On the other hand, from the standpoint of the solar empirical relation \citep[$d\Phi_{\rm e}/dt \propto \Phi^{0.3}$; ][]{2017ApJ...842....3N}, some corrections of the Otsuji's scaling law are necessary to theoretically understand the small power-law index.
%It may be better to reconsider their assumption (1) and (2).
%\textcolor{purple}{There is little support for the validity of the assumptions (1) and (2).}
For example, if emerging velocity has a negative dependence on the total flux (i.e., $v\propto \Phi^{\rm a}$, a $<$ 0), the empirical relation could be theoretically explained.
One explanation for the negative dependence is that the flux emergence could be suppressed to some extent if the surfaces are already filled with relatively strong magnetic fields.
As another hypothesis, since the emergence of a weak flux tube can be affected by convective motions \citep{2011ApJ...741...11W}, emergence velocity of small weak flux becomes faster if the field strengths have a positive relation with the total fluxes.

%The comparison with realistic numerical simulations can help us interpret the observations.
%As in Figure \ref{fig:emerge}, most of the emergence rates in numerical simulations take relatively higher values than solar observations while they looks roughly consistent with Otsuji's scaling law.
%Recent super-flaring AR12673 also shows extraordinary values 1.12$\times10^{21}$ Mx $\rm h^{-1}$ \citep{2017RNAAS...1a..24S} , which is comparable to the values of simulations $6\times 10^{20}$ $\rm Mx\cdot h^{-1}$ \citep{2014ApJ...785...90R}.
%However, it should be noted that the relatively high values may be due to the numerical set up like an initially given strong shear of flux.
%Since the numerical box and timescale are limited and they contain some special initial and boundary conditions, the straightforward comparison should be treated carefully.

As discussed above, the flux emergence process can be universally explained over a wide range of spot sizes including sunspots and star spots ($10^{18\sim24}$ Mx, $0.02\sim20000$ MSH), although the detailed understanding is not enough.
It should be noted that our results may contain some uncertainties and they can be updated by further studies. 
The emergence process of sunspots inside the convection zone is not well understood observationally even by the local helioseismology.
Although a comparison with numerical simulations can help us interpret the observations \citep[e.g.,][]{2014ApJ...785...90R}, the realistic ones with deep convection zone have not been done.
More researches on spot emergence by sunspot observations and simulations are necessary.

%%Simulation

%%Understandings of Starspots

%%Decay%%
\subsection{Decay of Star Spots}\label{sec:dis:decay}
Our results show that the decay rates of star spots are consistent with those of sunspots, and the sunspot distributions ($d\Phi_{\rm d}/dt \propto \Phi^{\sim0.49}$) almost correspond to the parabolic decay law \citep[$d\Phi_{\rm d}/dt \propto \Phi^{\sim0.5}$,][]{1993A&A...274..521M}.
This may suggest that sunspot and star spots are universally explained by the parabolic decay model, where spots decay by the erosion of the spot boundaries.

%\textcolor{blue}{Note that a small fraction of star spots are out of the extrapolated 95\% confidence intervals of sunspots which roughly corresponds to the parabolic decay law, while the remaining large fraction are included.}
%\textcolor{purple}{This may be (1) due to the rotational period dependence as mentioned in Section \ref{sec:res}, or (2) because consecutive spot emergence can occur even in the decay phase in the same region or reverse hemisphere.}

It should be noted here that \cite{1997SoPh..176..249P} suggested a corrected parabolic decay law in the form of $dA/dt \propto (A/A_{\rm max})^{0.5}$.
This may not match our data because the theory predicts that $dA/dt|_{A=A_{\rm max}}$ is a constant value independent of the spot size.

%In the case of the sunspots, most of the large sunspots consists of the several small sunspots, which can result in an apparent large power-law index in flux-decay rate relations.
%This can be considerable in the case of spatially unresolved star spot as well.
On the other hand, if we want to justify the linear decay theory ($dA/dt \propto$ constant), some excuses may be necessary.
In this case, decay rates should be, possibly apparently, enhanced only for large spots to explain the observations. 
Possible interpretations can be explained as follows: 
(1) The decay rates of large spots are apparently enhanced because large sunspots can consist of many small sunspots \citep[e.g.,][]{2008SoPh..250..269H} or different active regions on opposite latitudes.
In addition, many of large sunspots are classified into complex shapes, which can enhance flux cancellations \citep{1993A&A...274..521M}, although surface flux transport simulations are unsupportive for this \citep{2007A&A...464.1049I}.
(2) Stellar differential rotation can also contribute to high decay rate of large spots \citep{1994IAPPP..55...51H}.
\cite{2007A&A...464.1049I} also showed that differential rotations can accelerate flux cancelations depending on the tilt angle of bipole. 
(3) \cite{2014ApJ...795...79B} tried to explain the high magnetic diffusivity on stellar surface by assuming that supergranular scales determine the decay timescales, though the roles of supergranule are not well understood.

Recent MHD numerical simulations can give us more insights into the star spot decay.
\cite{2014ApJ...785...90R} performed 3-dimensional numerical simulations of spot emergence and decay.
The decay rate obtained by their simulations is also plotted in this Figure \ref{fig:decay}.
In the first decay phase in their simulations, the dispersal of flux is mainly due to the downward vertical convection motion.
In the late phase, intrusions of plasma in subsurface accelerate the spot fragmentations.
Their result indicates that the subsurface large-scale convectional flows can play a significant role in spot decay.
However, numerical box is limited only for the surface of convection zone and the subsurface morphology is little known observationally \citep{2011LRSP....8....3R}.
Further development of numerical simulations including a deep convection zone may reveal the spot decay mechanism.

Usage of the Doppler imaging technique can let us measure the temporal evolutions of star spots as well.
%Because the continuous high-dispersion spectroscopic observation through one year is quite expensive and the Doppler imaging of slowly-rotating solar-type stars is quite difficult, a decay rate of a star spot on a solar-type star has not been measured.  
A decay rate of the red giant star XX Triangulum was estimated to be -5.6$\times 10^{22}$ Mx$\cdot$h$^{-1}$ (-920 MSH$\cdot$h$^{-1}$) with the area of 1.2$\sim$6.2$\times 10^{26}$ Mx (2$\sim$10$\times10^{6}$ MSH) according to the Doppler imaging method \citep{2015A&A...578A.101K}.
The decay rate is surprisingly consistent with the parabolic decay relation in our paper (Figure \ref{fig:decay}), although its surface effective temperature (4,620 K) and gravity (log $g$ = 2.82) are quite different from those of solar-type stars.
Red giant stars or cooler stars will be subjected to consideration in our next paper, as these stars are currently beyond the scope of this paper.

This is an incipient research on the temporal evolutions of unresolved star spots.
%As mentioned by \cite{2014ApJ...785...90R}, penumbra can play a significant role in spot stabilization and destabilization, and penumbral area is completely unknown in the star spot.
%For these reason, one must be carful in interpreting the comparisons between sunspot and star spot data.
Further development in sunspot and star spots observations as well as numerical simulations are required for the understandings of decay process of star spots.

%%Lifetime%%
\subsection{Lifetime of Star Spots}

Historically, the spot evolutions have been discussed in terms of the lifetimes, and the simple GW law ($T \propto A$) has been used in the solar community (see Section \ref{sec:int}).
In contrast, the lifetimes (10 days $\sim$ 350 days) of star spots on solar-type stars are much shorter than those extrapolated from the GW relation (300 days $\sim$ 1000 days).

%The dependence of emergence and decay rates on the spot area can help us to understand this trend.
If we assume that the emergence and decay rates are independent of the (total) spot fluxes ($d\Phi_{\rm e, d}/dt$ = constant), spot lifetimes naturally follow the GW law \citep{1974MNRAS.169...35M}, which is inconsistent with star spot data.
However, as discussed in Section \ref{sec:dis:emerge} and \ref{sec:dis:decay}, the emergence and decay rates clearly depend on the spot area across wide ranges of total fluxes ($d\Phi_{\rm e, d}/dt \propto \Phi^{0.3\sim 0.5}$).
This dependence of variation rates on the maximum sizes would be one reason why the lifetimes of star spots are much less than those expected from the GW law.
In this case, the relation between lifetimes and area can be expressed as $T\propto A^{0.5\sim 0.7}$.
We discuss detection limits and method dependences on star spot lifetimes in Section \ref{sec:md}.

Interestingly, ground based observations have revealed that HR 7275, a RS CVn-type K1 III-IV star, has spots with much longer lifetimes \citep[$\sim$2.2 years on average,][]{1994A&A...282..535S}, although the spot areas are also huge.
Moreover, cooler stars are reported to have longer lifetimes \citep{2017MNRAS.472.1618G}.
The differences in lifetimes can reveal the role of surface convection in spot decay.
Moreover, in the rapidly rotating young star V410 Tau, a large spot near the pole has persisted for at least 20 yr \citep{2012A&A...548A..95C}, and FK Comae giant HD 199178 has a polar spot whose lifetime is more than 12 yr \citep{1999A&A...347..212S}.
These properties of polar spots can be different from the solar-like spots at lower latitude.
The comparison between different types of stars are beyond the scope of this paper, but will be addressed in future.

%Moreover, the expected small diffusion coefficient of stellar surface can affect the numerical simulations of stellar activity cycles as well \citep{2014ApJ...795...79B}.

%\textcolor{purple}{Note that the GW relations are indicated to be also derived by the parabolic decay law \citep{1997SoPh..176..249P}, so that the above discussions may not be incomplete.}

%As discussed above, lifetimes can provide information on the mechanism of temporal evolutions of star spots.

\subsection{Rotational Period Dependence}

Since the rotational period is a good indicator of the stellar ages, the dependence on the temporal evolutions can be a hint to the understandings of the evolutions of stellar dynamo.
As in Section \ref{sec:res}, the star spots on rapidly rotating stars tend to show more rapid emergences and decays compared to the spots on the slowly rotating ones.
The rapid decay can be explained by the stellar differential rotation because rapidly rotating stars are thought to have a strong differential rotations \citep[e.g.,][]{2011ApJ...740...12H,2016MNRAS.461..497B}.
Here, we only selected star spot pairs whose relative longitude are considered to be unaffected by the stellar differential rotation.
Since the rotational periods of the pairs are not completely the same value, there is a possibility that the strong surface differential rotations on rapidly rotating stars makes it difficult to trace the local minima for a long time, which can result in short lifetimes in rapidly rotating stars.
On the other hand, according to \cite{2017MNRAS.472.1618G}, there is no clear dependence on the rotational period on the decay timescale of star spots.
They however analyzed only the relatively slowly rotating stars (10 and 20 days), and the application to the more rapidly rotating stars have not yet been done.
The detailed dependence of rotational periods should be researched in future.

\subsection{Uncertainties on the Measurements of Lifetime and Area}\label{sec:md}

Let us summarize here uncertainties and biases of the results which can be caused by our method.
First, we do not correct the star spot area for the stellar inclination angles.
The star spot area, as well as variation rates, can be somewhat underestimated if the stellar inclination angles (sin \textit{i}) are small \citep{2015PASJ...67...33N} or the spots are located on the high latitude.
Under the assumption of the random orientation of rotational axes, the average inclination angle ($i$) can be derived as 1 radian.
If we assume this typical inclination angle, the stellar inclination reduces the statistical values of star spot area by $\sim$30\% for the sun-like star spot distribution \citep[latitude $\sim$ 10-30$^{\circ}$;][]{2003A&ARv..11..153S}.

Second, the determination of the unspotted brightness levels of the stars is subjected to difficulty, affected by the existence of faculae and large polar spots, and the inevitable Kepler's long-term observational trend.
However, we calculate the spot area from the brightness differences from the local maxima and minima.
These are the relative values and less affected by the zero-levels.
Even if the stellar unspotted level becomes brighter by 1\% \citep[e.g. solar faculae;][]{2013ARA&A..51..311S}, the spot size can decrease only by a few per cent. 
Likewise, we ignore the contributions of stellar faculae to the stellar brightness variations because the distributions and filling factors are not well known.
If the faculae are localized in a single hemisphere and the brightness contributions of faculae are comparable to those of spots, the spot area can be both over- and under-estimated to some extent.

Third, since we used the local depth as the spot area, contaminations of other spots are not corrected in this analysis. 
This also contribute the uncertainties of the estimation of spot area.
To avoid this effect, light-curve modelings with several star spots would be necessary for more detailed analyses.

Moreover, it should be noted here that lifetimes can have some uncertainties due to observational and analytical problems.
The star spots become difficult to detect as the spot area decrease depending on the photometric errors and analysis methods, which leads to the underestimations of the spot lifetimes.
Although we extrapolated the lifetimes by assuming linear emergence and decay, it is just an assumption.
In addition, there can be a selection bias that long-lived spots with lifetimes of $\gtrsim$1,000 days can be missed because the Kepler observational period is limited to only $\sim$4 years.
In this case, the emergence and decay rates can be smaller than our results.

\cite{2014ApJ...795...79B} and \cite{2015PhDT.......177D} have reported the lifetime of star spots by using exoplanet transits, and their results are somewhat different from our results.
Their methods have an advantage that they can spatially resolve the stellar surface, and their lifetime and star spot area may more clearly represent the single star spot properties.
Note that the estimated area are different by a factor of $\sim$3 between the result of \cite{2014ApJ...795...79B} and \cite{2015PhDT.......177D}, though they analyze the same star (Kepler-17).
%This discrepancy indicates that the estimations of spot area can depend on the analysis methods.
%The result of our study and \cite{2017MNRAS.472.1618G} are also inconsistent with those of \cite{2014ApJ...795...79B} and \cite{2015PhDT.......177D}.
%Moreover, the Kepler-17 is a solar-type star, but the system has a hot-jupiter planet ($M\sim 2.5M_{\rm jupiter}$) at 0.03AU from the central star \citep{2011ApJS..197...14D}.
%There is a possibility that a hot jupiters can affect the spot activity of the central star \citep[e.g.,][]{2004ApJ...602L..53I,2013PASJ...65...49S}, though the details have not been confirmed.

\section{Summary and Conclusion}

Many solar-type stars show extraordinary high magnetic activities, which cannot be expected from the solar observations, such as spot activity and superflares.
The large star spots are considered to be a key to understand the superflare events as well as underlying stellar dynamo.
The subject of this study is to investigate the emergence and decay processes of large star spots on solar-type stars by comparing them with well-known sunspot properties.
We have developed a method to measure the temporal evolutions of single star spots by tracing local minima of visible continuum brightness variations.
By applying this method to a huge amount of \textit{Kepler} data, we have successfully detected temporal evolutions of star spots showing clear emergence and decay phase.

We mainly obtained the following three results: 
(i) the emergence rates of star spots are consistent with those extrapolated from the sunspots ($d\Phi_{\rm e}/dt \propto \Phi^{0.3\sim 0.5}$) under the assumption of the spot magnetic field strengths of 2000 G;
(ii) the decay rates of star spots are consistent with those extrapolated from the sunspots ($d\Phi_{\rm e}/dt \propto \Phi^{\sim 0.5}$), which might be understood as an erosion from the edge;
(iii) the lifetime of star spots are much shorter than those extrapolated from the empirical GW relation of sunspots ($T \propto A$), though the lifetimes are up to 1 year.
The results (i) and (ii) indicate that emergence and decay of sunspots ($10^{18-23}$Mx, 0.02-2000 MSH) and large star spots on solar-type stars ($10^{23-24}$Mx, 2000-20000 MSH) can be universally explained by the same underlying process, i.e. a flux emergence from stellar interior and a following flux diffusion in stellar surface.
Lifetimes have been used as an indicator of spot temporal evolutions, but comparisons with sunspot lifetimes should be carefully made because the lifetimes of star spots can be underestimated due to the data sensitivities.
Nevertheless, the large star spots ($\sim$10,000 MSH) potential to produce superflares \citep[$\sim 10^{34}$ erg,][]{2013PASJ...65...49S} are found to survive for up to $\sim$1 year.
This implies that the surrounding exoplanets can be exposed to danger of superflares for such a long time.
Moreover, according to frequency distributions of superflares, superflares of $10^{34}$ erg can occur about once a year on the star spots with area $\sim$10,000 MSH \citep{2017PASJ...69...41M}.
This may indicate that supreflares can occur with an high probability once such large spots emerge on the stellar surfaces.

In the light curves among our spot candidates, there are some transient brightness enhancements which are probably superflares (see online figures).
In the case of the solar flares, it is known that many of the strong flares are caused by newly emerging flux adjacent to the pre-existing sunspots \citep{1987SoPh..113..267Z,2017ApJ...834...56T}.
As future works, the timing of superflares in the spot temporal evolutions should be investigated to understand how the superflares are triggered.
Moreover, in solar case, complex sunspots are considered to have high magnetic free energy \citep{2017ApJ...850...39T}, showing high flare occurrence rates \citep{2000ApJ...540..583S,2017PASJ...69...41M} and short lifetimes \citep{1993A&A...274..521M}.
This implies that flare occurrence rates and lifetimes can become an indicator of spot configurations \citep[e.g.,][]{2017PASJ...69...41M}.

Finally, we also found that star spots on rapidly rotating stars show more rapid temporal evolutions than those on slowly rotating stars. 
%we did not focus on the dependences of stellar properties ($T_{\rm eff}$ and $P_{\rm rot}$) on the temporal evolutions of star spots.
These dependence on the rotational period as well as stellar effective temperature should be addressed in future works.
Also, we have not examine some uncertainties described in Section \ref{sec:md}.
It is necessary to perform the spot modelings \citep[e.g.][]{2012A&A...543A.146F}, inversion modelings \citep[e.g.][]{2008AN....329..364S}, and follow-up spectroscopic observations for the further evaluations.

\bigskip
Acknowledgement: We acknowledge with thanks S. Takasao, H. Hotta, M. Kubo, A. Norton, K. Petrovay, M. G\"udel and T. L\"uftinger for their fruitful comments on our work. 
We also thank D. Hathaway, K. Petrovay, L. van Driel-Gesztelyi, K. Otsuji and T. A. Berger for kindly providing us their data.
Kepler was selected as the tenth Discovery mission.
Funding for this mission is provided by the NASA Science Mission Directorate.
The data presented in this paper were obtained from the Multimission Archive at STScI.
The SORCE program is supported under NASA contract NAS5-97045 to the University of Colorado. 
This research was supported by a grant from the Hayakawa Satio Fund awarded by the Astronomical Society of Japan.
This work was also supported by JSPS KAKENHI Grant Numbers
JP15H05814, % (Toriumi, PSTEP, A02)
JP16H03955, % (Shibata), 
JP16J00320, % (Notsu Y.),
JP16J06887, % (Notsu S.),
JP16K17671, % (Toriumi)
JP17H02865, % (Nogami),
JP17K05400, % (Maehare),
JP17J06954, % (Hayakawa),
and JP18J20048% (Namekata)

%\setcounter{figure}{0}
%\begin{figure}[htbp]
%\begin{center}
%\includegraphics[scale=0.35]{figures/lifetime-02-11137355.ps}
%\includegraphics[scale=0.35]{figures/lifetime-03-12314597.ps}
%\includegraphics[scale=0.35]{figures/lifetime-04-1865674.ps}
%\includegraphics[scale=0.35]{figures/lifetime-05-3110573.ps}
%\end{center}
%\caption{Continued.}
%\label{fig:oc1}
%\end{figure}
%\clearpage

\appendix

\section{Validation of Method: The Sun-as-a-star Analysis}\label{app1}
\begin{figure}[htbp]
\begin{center}
\includegraphics[scale=0.6]{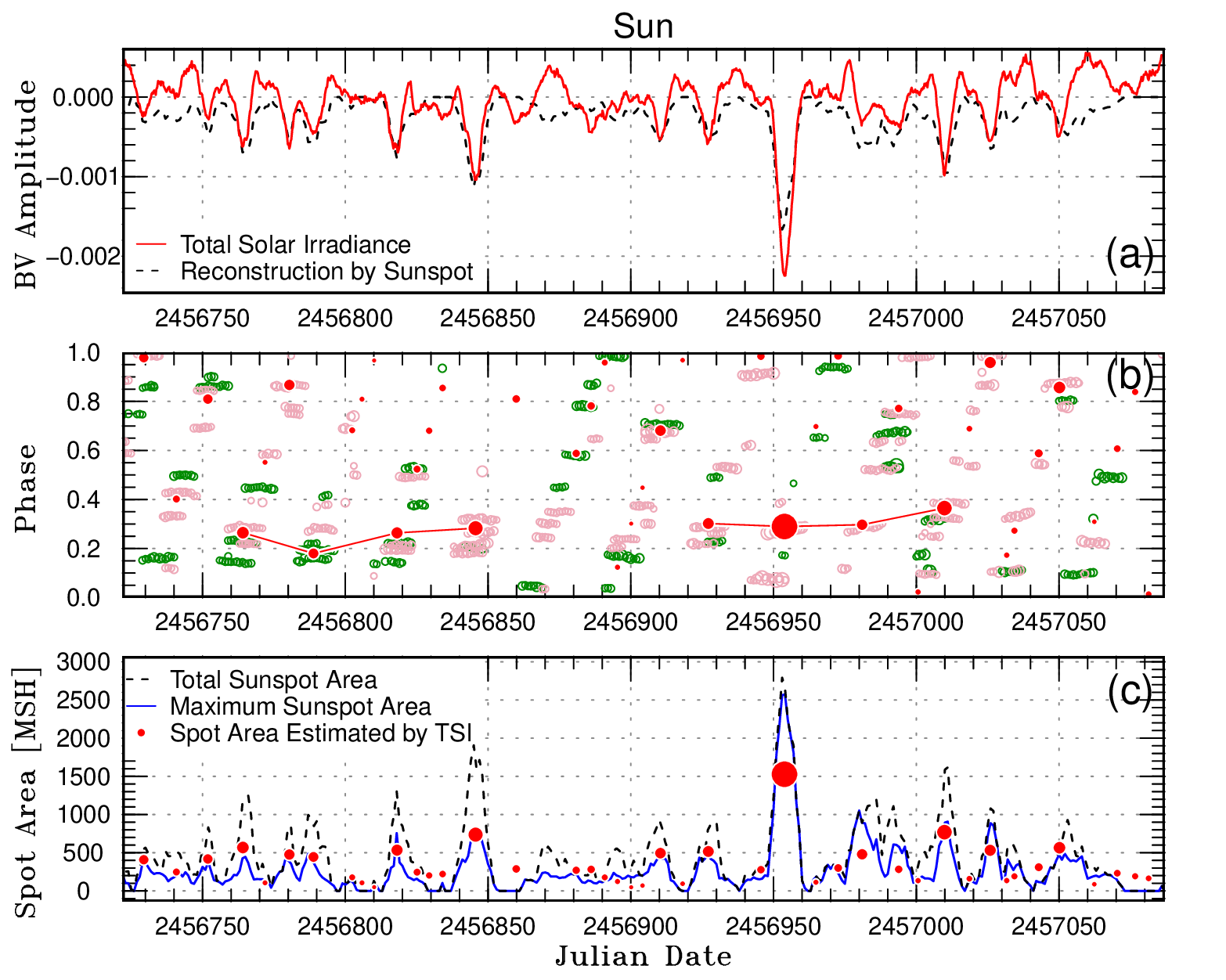}
\caption{An example of the sun-as-a-star analysis for from 5th March 2014 to 5th March 2015.
(a): A red solid line is the light curve of total solar irradiance (TSI) observed by the \textit{SORCE} satellite.
The data are plotted with 6 hours time cadence, and the values are normalized by the median values of the TSI in the observational period. 
A black dashed line is also the light curve reconstructed from the daily sunspot area.
(b): Phase-time diagram.
Red points correspond to the local minima detected in the upper panels compared to the rotational period (cf., Carrington longitude).
Background green and pink open circles are the spatially resolved sunspot locations in northern and southern solar hemisphere, respectively.
In this panel, sunspot data are plotted only for sunspots with the area $>$ 100 MSH.
Symbol sizes correspond to the spot area, but the scale of red and others are different for visibility.
(c): Temporal evolution of sunspot area.
Red circles correspond to the area estimated from the depth of the local minima of TSI.
The dashed black line represents the daily total sunspot area, and the blue solid line is the area of largest sunspot for each date.}
\label{app2:sun}
\end{center}
\end{figure}

\begin{figure}[htbp]
\begin{center}
\includegraphics[scale=0.6]{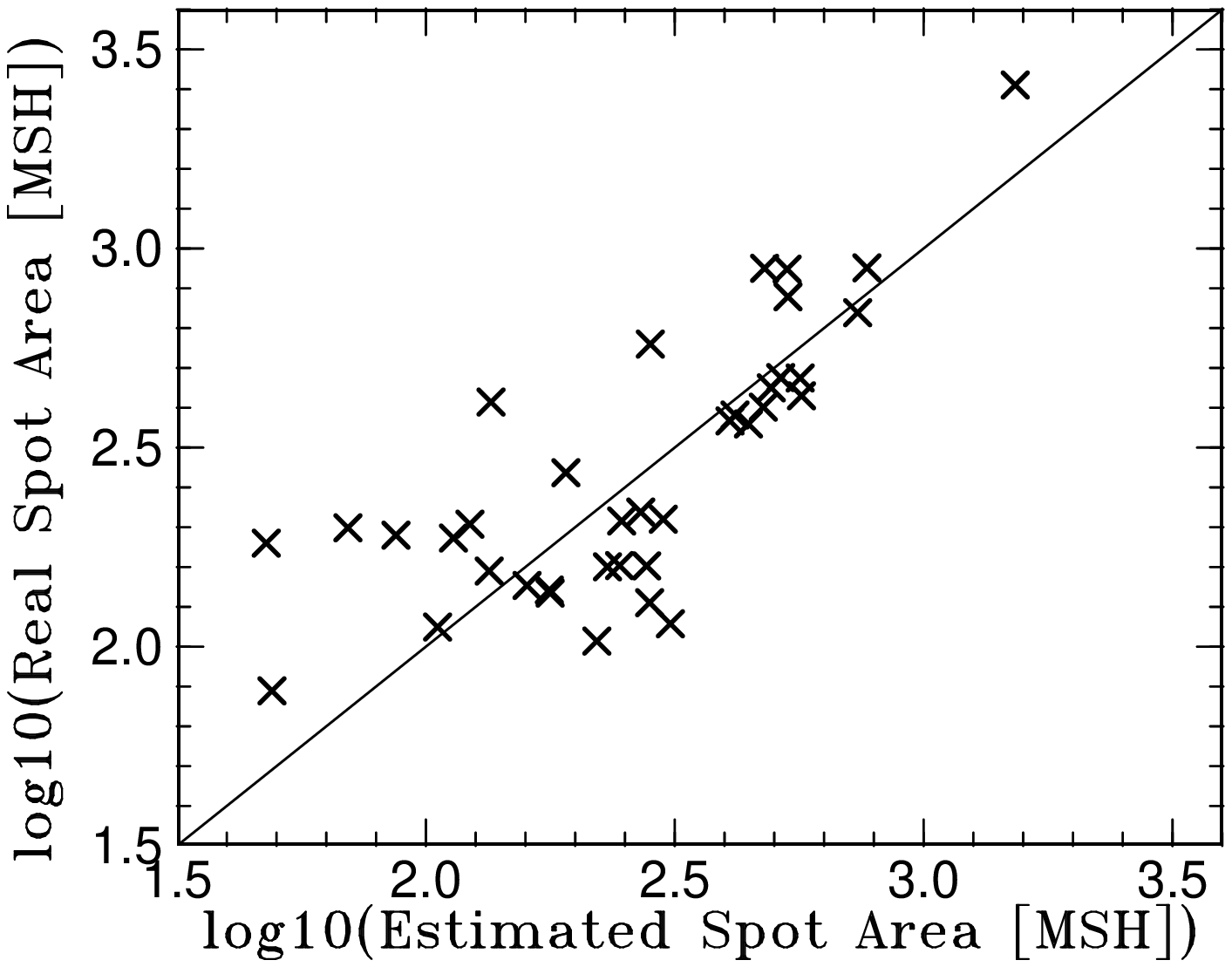}
\caption{Comparison between the estimated spot area from the TSI and the real values in Figure \ref{app2:sun}(c). The data are plotted for MJD 2456722 to MJD 2457072}
\label{app2:sunasastar}
\end{center}
\end{figure}

To evaluate the sensitivities of the estimation of spot area, we simply tested our method using the solar data since we can spatially resolve the sunspot distributions.
We used the total solar irradiance (TSI) observed by Total Irradiance Monitor onboard the \textit{SORCE} satellite \citep{2005SoPh..230....7R}, as a proxy of the \textit{Kepler} light curve. 
The TSI is the spatially and spectrally integrated absolute intensity of solar radiation, at the top of the Earth atmosphere.
We use the 6-hours averaged time series from 5th March 2014 to 5th March 2015.
After removing high frequency component of the light curve, we detected the local minima, measured the local depth of the local minima, and estimated the spot area in the same manner of our star spot analysis.
Figure \ref{app2:sun} shows the comparison between the Sun-as-a-star analysis (estimation from TSI) and the analysis based on spatially resolved sunspot data (answer from sunspot observations).
The sunspot data are based on observations by The Royal Greenwich Observatory (RGO) sunspot data (https://solarscience.msfc.nasa.gov/greenwch.shtml).
The TSI well match the light curve reconstructed from sunspot distribution by assuming a simple limb darkening ($I(\theta)/I(0)=1-0.6(1-\rm cos \it \theta)$, see \cite{2013ApJ...771..127N}).
The estimated spot area are roughly consistent with the actual values (lines) in Figure \ref{app2:sun}(c).
Figure \ref{app2:sunasastar} shows a good agreement between the estimated spot area from the TSI and the real values.
One reason of the little difference would be due to the the projection effect because sunspot are generally located on latitude of 20-30$^\circ$, which reduces the area by a factor of 0.9.
The brightness variation by faculae regions can also contribute to the errors, and lack of limb-darkening calculation can cause the underestimation.
Interestingly, in Figure \ref{app2:sun}(a), we can see the anti-correlation variation between black and red around JD 2456870.
The time corresponds to one rotational period after the large sunspot group has disappeared (the red line in the middle panel).
This may be due to the faculae whose magnetic flux has diffused from the decaying sunspot.
Here we demonstrated that our method can simply estimate the spot area and lifetime from the integrated light curve, though brightness variation of faculae can provide some errors.

\end{document}